\pdfoutput=1

\documentclass[11pt]{article}
\usepackage{graphicx}
\graphicspath{{./figures/}}
\usepackage{amsmath,amssymb,bm}
\usepackage{siunitx}
\usepackage{booktabs,array}
\usepackage{braket}
\usepackage[super,sort&compress]{natbib}
\setcitestyle{citesep={,}}
\usepackage{xr-hyper}
\usepackage{url}
\usepackage[capitalize, nameinlink]{cleveref}
\DeclareSIUnit{\dB}{\deci\bel}

\newcommand{\Real}{\operatorname{Re}}
\newcommand{\Imag}{\operatorname{Im}}
\newcommand{\Tr}{\operatorname{Tr}}
\newcommand{\Fnorm}{F_C^{\rm norm}}
\newcommand{\Fmean}{F_C^{\rm mean}}
\newcommand{\Fcov}{F_C^{\rm cov}}
\newcommand{\Gfwd}{G_{\rm fwd}}
\newcommand{\Grev}{G_{\rm rev}}
\newcommand{\Iso}{\mathcal{I}}

\newcommand{\share}{\rho}

\crefname{equation}{Eq.}{Eqs.}
\Crefname{equation}{Equation}{Equations}
\crefname{table}{Table}{Tables}
\crefname{figure}{Fig.}{Figs.}
\Crefname{figure}{Figure}{Figures}
\crefname{section}{Sec.}{Secs.}
\crefname{subsection}{Sec.}{Secs.}

\usepackage[margin=2cm]{geometry}
\title{Forward gain tracks the mean-response share of the Fisher information in non-Hermitian bosonic ladders}
\author{Juste Deuyekbe$^{1}$, P.~Djorw\'e$^{1,2}$, Nsangou Mama$^{3}$, S.~G.~Nana Engo$^{4,*}$}
\date{%
\small $^{1}$Department of Physics, Faculty of Science, University of Ngaound\'er\'e, P.O.~Box 454, Ngaound\'er\'e, Cameroon\\
\small $^{2}$Stellenbosch Institute for Advanced Study (STIAS), Wallenberg Research Centre, Stellenbosch University, Stellenbosch 7600, South Africa\\
\small $^{3}$Department of Physics, Faculty of Science, University of Maroua, P.O.~Box 46, Maroua, Cameroon\\
\small $^{4}$Department of Physics, Faculty of Science, University of Yaound\'e~I, P.O.~Box 812, Yaound\'e, Cameroon\\
\small $^{*}$serge.nana-engo@facsciences-uy1.cm}

\begin{document}

\maketitle

\begin{abstract}
Non-Hermitian lattices are proposed as sensors because non-normal amplification turns a small frequency shift into a large output, yet whether this increases precision at fixed photon number remains contested. We evaluate the classical Gaussian-state Fisher information of the intracavity field, split into mean-response and covariance-response terms, for a single bosonic chain and a two-chain opposing-skin ladder across \num{34560} points under a Hurwitz stability margin. Over the \num{15432} configurations passing both the margin and an independent covariance check, the mean-response share of the clean-cavity benchmark tracks forward gain along one curve in both architectures ($R^2=\num{0.978}$; architecture increment $\Delta R^2=\num{0.0012}$), falling from \num{0.97} below zero gain to below \num{2e-3} above \qty{60}{\dB}. Inter-chain coupling shifts the accessible gain range without raising the share at matched gain. Amplification thus relocates information into an ill-conditioned noise covariance, making the mean--covariance split and covariance conditioning the decisive metrics.
\end{abstract}


\section{Introduction}

Non-Hermitian lattices concentrate their eigenstates at a boundary, and the resulting non-normal amplification has been proposed as a sensing resource: a frequency perturbation at one site emerges at another multiplied by an exponential of the lattice length \citep{Hatano1997,Yao2018,Kunst2018,Bergholtz2021,Weidemann2020,Helbig2020}. Exceptional points sharpen the same premise, yielding a spectral splitting that scales as a fractional power of the perturbation \citep{Wiersig2020,Hodaei2017,MiriAlu2019,ElGanainy2018,Ozdemir2019,Ashida2020}. Whether this amplified response delivers genuinely higher precision at a fixed photon budget remains the central open question.

That promise has faced sustained theoretical scrutiny. Non-normal amplification multiplies signal and noise together, so gain-based figures of merit overstate metrological performance once intracavity photon number is held fixed \citep{Lau2018,McDonald2022,McDonald2020Exponential,Bao2021,Ding2023}. The Petermann factor penalizes the same eigenvector non-orthogonality that generates the gain \citep{Wang2020Petermann}, and proximity to an exceptional point does not outpace realistic measurement noise \citep{Langbein2018,Duggan2022}. Recent analyses have sharpened these constraints along three measurement classes. They bound sensitivity by the spectral width of the signal-carrying mode \citep{Liu2026SpectralWidth,Wiersig2026,LiGuo2026,Carney2026}, track information under post-selected detection \citep{Zeng2025PostSelected,Arkhipov2026PseudoHermitian,Practical2026}, and formulate linear response through Green functions \citep{Liang2026,XieXu2024}. Each addresses a distinct operational object—a spectral-width ceiling, a post-selected rate, or a linear susceptibility. Establishing an architecture-level design rule requires comparing candidate lattices directly at matched resource normalization, matched stability margin, and matched forward gain.

Coupled non-Hermitian chains offer a natural candidate for separating directional signal amplification from noise accumulation \citep{Metelmann2015,McDonald2020Exponential}. In the minimal geometry—a two-chain bosonic ladder with opposing non-reciprocal hoppings—the forward channel of one chain runs opposite to that of the other, so excitations crossing between chains encounter an opposing skin accumulation. Here we test whether this geometry improves fixed-photon precision inside the Hurwitz-stable regime using differentiable optimal control and a matched parameter-box comparison.

Evaluating the full classical Gaussian-state Fisher information \citep{Pinel2013,Safranek2019} of the linearized intracavity steady state yields both a physical design rule and a methodological benchmark. Physically, resource-normalized Fisher information does not saturate at unity; its apparent divergence near the stability threshold is carried by the frequency derivative of the noise covariance rather than by the mean field, and terminates where covariance ill-conditioning exhausts double-precision arithmetic. Separating the mean-response and covariance-response terms reveals a compact scaling law: across both architectures and all tested lattice sizes, the share of the clean-cavity benchmark retained in the mean field collapses onto a single curve in forward gain, governed algebraically by the Rayleigh quotient of the inverse covariance along the amplified mode. Inter-chain coupling shifts where along that curve the ladder can remain stable—via a closed-loop opposing-skin feedback instability—without raising the mean-response share at matched gain.

Methodologically, realifying a complex multi-mode Gaussian state to solve its real Lyapunov equation requires aligning the read-out derivative with the block offset of the realification rather than an interleaved site stride. The two conventions coincide for a single resonator ($N=1$) but diverge for arrays ($N\ge2$), where an interleaved index suppresses the amplifying branch and creates the false appearance of a unimodal ceiling near unity.

\section{Results}

\subsection{Model, estimator and the clean-cavity reference}

\begin{figure}[t]
\centering
\includegraphics[width=0.92\textwidth]{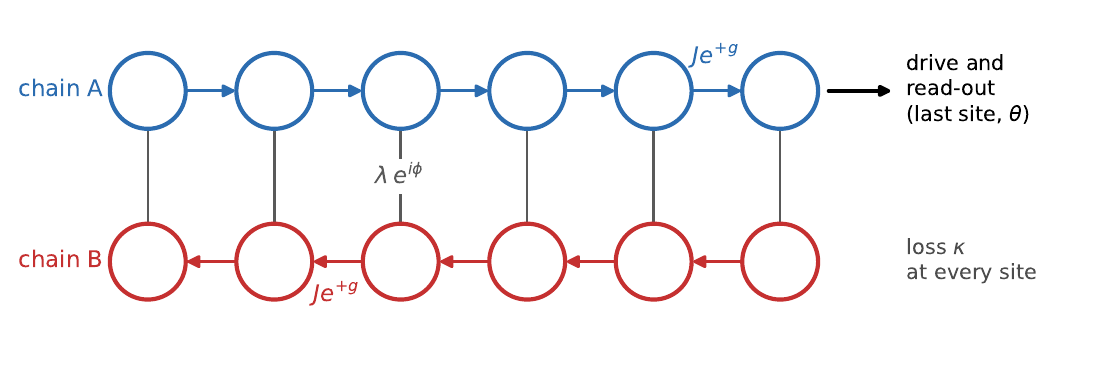}
\caption{The opposing-skin bosonic ladder. Two chains of $N$ sites carry opposite non-reciprocal nearest-neighbor hoppings: $J e^{+g}$ rightward and $J e^{-g}$ leftward in chain A, reversed ($J e^{+g}$ leftward and $J e^{-g}$ rightward) in chain B, coupled site by site with amplitude $\lambda$ and phase $\phi$. Every site loses photons at rate $\kappa$. The frequency shift $\theta$ acts on the last site of chain A, which serves as the driven port.}
\label{fig:schematic}
\end{figure}

Each chain has $N$ sites with nearest-neighbor hopping $J e^{\pm g}$, the sign alternating between the chains so that their skin accumulations oppose. The two chains are coupled site by site with amplitude $\lambda$ and phase $\phi$. Every site loses photons at amplitude rate $\kappa$, the signal is a frequency shift $\theta$ applied at the last site of chain A, and that site is the driven port. \Cref{fig:schematic} shows the geometry. In the linearized Gaussian regime the dynamics reduce to a complex drift matrix $A(\theta)$ of dimension $2N$, a drive vector $u$, and a diffusion matrix $2\kappa I$ acting on both chains; the steady state is Gaussian, with mean $\mu = -A^{-1}u$ and Hermitian covariance solving $A\Sigma + \Sigma A^{\dagger} = -2\kappa I$ (realified into $A_{\rm R}\Sigma_{\rm R}+\Sigma_{\rm R}A_{\rm R}^{\mathsf T}=-2\kappa I$ in~\cref{sec_realification}).

The classical Fisher information that this Gaussian steady state carries about $\theta$ splits into two terms,
\begin{equation}
F_C \;=\; \underbrace{\dot\mu_{\rm R}^{\mathsf T}\Sigma_{\rm R}^{-1}\dot\mu_{\rm R}}_{\Fmean} \;+\; \underbrace{\tfrac12\Tr\!\left[\left(\Sigma_{\rm R}^{-1}\dot\Sigma_{\rm R}\right)^{2}\right]}_{\Fcov},
\label{eq:split}
\end{equation}
where the dot denotes $\mathrm d/\mathrm d\theta$ and we drop the subscript ${\rm R}$ on $\mu$ and $\Sigma$ below whenever realified quantities are meant. Equation~\eqref{eq:split} is the classical Fisher information of the intracavity Gaussian distribution in units where the vacuum noise is one, rather than the full quantum Fisher information (QFI) of a Gaussian state \citep{Pinel2013,Safranek2019}. The first term $\Fmean$ coincides in form with the displacement contribution of the Gaussian QFI and quantifies the information accessed by a linear displacement measurement of the field. The second term $\Fcov$ quantifies the classical sensitivity of the intracavity noise covariance to $\theta$; unlike the displacement term, it vanishes unless the covariance itself depends on $\theta$, and phase-insensitive added noise dilutes it least where it already dominates. The intracavity field is not propagated to an output detector in~\cref{eq:split}; port transmission and detector noise enter separately through the scattering analysis of Supplementary \cref{SI-sec:si_platform}.

Resource normalization divides~\cref{eq:split} by the steady-state intracavity photon number $\|\mu\|^{2}$, summed over both chains, defining $\Fnorm = F_C/\|\mu\|^2$ (with dimensions of $\kappa^{-2}$ per photon). The benchmark of the analysis is a single cavity driven and read at the same port, with drift $A=-\kappa-i\theta$. In the vacuum-noise normalization its covariance is $\Sigma=I$ independently of $\theta$, so $\Fcov\equiv 0$ and
\begin{equation}
\kappa^2 \Fnorm \;=\; \frac{1}{1+(\theta/\kappa)^{2}},
\label{eq:cavity}
\end{equation}
exactly for every drive amplitude. At the operating detuning $\theta_0=0.1\kappa$ used throughout, \cref{eq:cavity} gives $\num{0.9901}$. We therefore define the \emph{mean-response share} of the clean-cavity benchmark,
\begin{equation}
\share \;=\; \frac{\kappa^2 \Fmean/\|\mu\|^{2}}{1/[1+(\theta_0/\kappa)^{2}]},
\label{eq:share}
\end{equation}
which measures what fraction of the clean-cavity response an architecture retains in its mean field. Because $\share$ is normalized to the clean-cavity benchmark rather than to $F_C$, it is distinct from $\Fmean/F_C$ and is not strictly bounded above by one: it slightly exceeds unity where the normalized mean response surpasses the bare cavity, reaching $\num{1.024}$ at $N=2$ in~\cref{tab:landscape} and $\num{1.0002}$ across the sampled configurations. Conversely, a configuration whose information resides in noise fluctuations has $\share\ll1$ however large $\Fnorm$ becomes.

\subsection{Realification indexing in multi-mode Gaussian metrology}

Evaluating~\cref{eq:split} via automatic differentiation turns the complex drift of dimension $m=2N$ into a real matrix of dimension $2m=4N$ whose covariance solves a real Lyapunov equation. In the block convention, which stacks the real and imaginary parts of the full state vector, site $j\in\{0,\dots,N-1\}$ of chain A occupies real index $j$ and its quadrature partner occupies index $m+j$. The read-out port $j=N-1$ therefore sits at index $N-1$, and the detuning generator acts on the pair $(N-1,\,3N-1)$. An implementation that instead assumes an interleaved two-quadrature stride $(2j,2j+1)$ places the read-out derivative at index $2(N-1)$, differentiating an uncoupled entry for $N\ge2$. Because both conventions map a single resonator ($m=N=1$) to the same index pair $(0,1)$, this stride mismatch passes single-cavity unit tests silently.

\begin{figure}[t]
\centering
\includegraphics[width=0.72\textwidth]{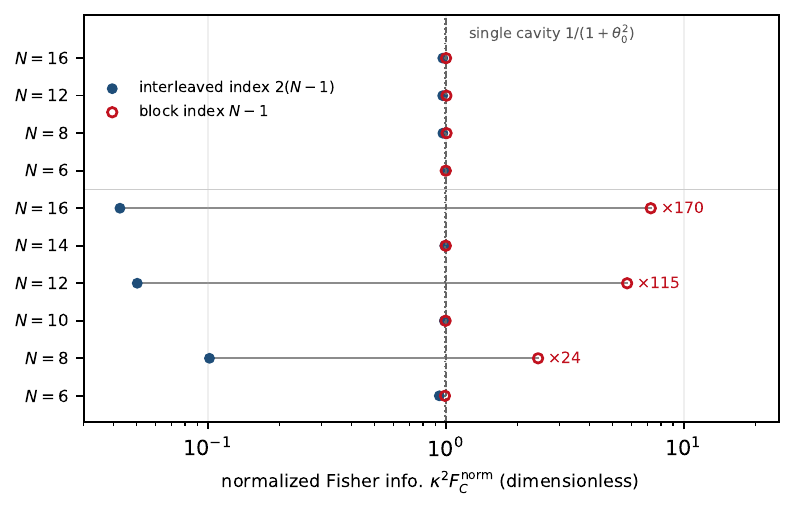}
\caption{Effect of realification indexing on $\kappa^2\Fnorm$ at ten representative ladder operating points, evaluated with an interleaved stride index $2(N-1)$ (filled circles) versus the block-aligned read-out index $N-1$ (open circles). Pairs are connected; ratios are printed where they differ by more than ten percent. The upper four points lie in the deep-stability regime (1500-step continuation) and the lower six in the amplifying benchmark set (500-step starts), distinct from the ten-restart optima of~\cref{tab:landscape} (which reach $\kappa^2\Fnorm=\num{67.2}$ at $N=8$). Dotted line: clean-cavity benchmark of~\cref{eq:cavity}; dashed line: unity.}
\label{fig:metric_audit}
\end{figure}

\Cref{fig:metric_audit} quantifies the resulting distortion across ten representative ladder operating points spanning deep-stability and amplifying regimes. While the four deep-stability points agree to within four percent, the interleaved stride suppresses the amplifying configurations by factors of \num{24} ($N=8$), \num{115} ($N=12$), and \num{170} ($N=16$). Specifically, misplacing the read-out index erases the amplifying branch of the optimization landscape, creating the false appearance of a unimodal bound $\kappa^2\Fnorm\le1$ saturating near the single-cavity value $\num{0.9901}$. Under block-aligned indexing, $\kappa^2\Fnorm$ exceeds unity across the amplifying branch, reaching \num{2.44} at $N=8$, \num{5.77} at $N=12$, and \num{7.25} at $N=16$ at the same stable operating points (Supplementary \cref{SI-sec:si_index}).

\subsection{The mean-response share tracks the forward gain}

We sampled one common parameter box for both architectures, with $J\in[0.05,0.5]\kappa$ (\num{32} values) and $g\in[0.5,3.0]$ (\num{30} values) throughout, and $\lambda\in[0.005,0.1]\kappa$ (\num{8} values) for the ladder, across $N=6$, $8$, $10$, and $12$ (\num{34560} parameter points in total: \num{3840} single-chain and \num{30720} ladder). Retaining only configurations that satisfy the Hurwitz stability margin $\max\Real\nu<-0.05\kappa$, carry finite positive information, and agree to within one percent in $\Fnorm$ against an independent Bartels--Stewart Schur solver \citep{Bartels1972} leaves \num{15432} validated configurations—all \num{12866} margin-stable ladder points together with \num{2566} single-chain points (full filtration breakdown in~\cref{sec_reliability}). Whereas this matched grid covers $N=6$--$12$, the optimal-control landscape of~\cref{tab:landscape} spans $N=1$--$16$.

\begin{figure}[t]
\centering
\includegraphics[width=\textwidth]{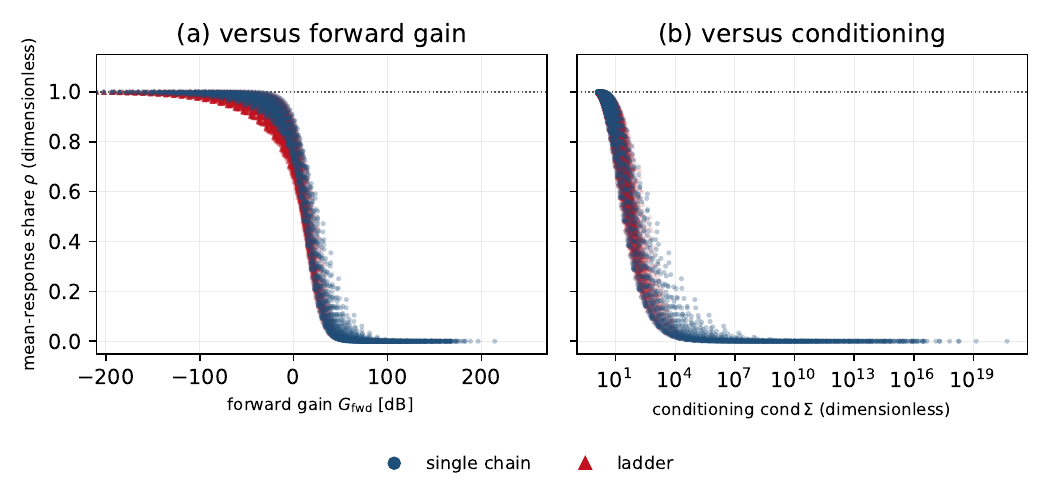}
\caption{Mean-response share~\cref{eq:share} for all \num{15432} retained configurations: single chain (blue circles, \num{2566} points) and opposing-skin ladder (red triangles, \num{12866} points) across $N=6$, $8$, $10$, and $12$. (a) Versus forward gain $\Gfwd$: both architectures and all four sizes follow one empirical trend ($R^2=\num{0.978}$, $\Delta R^2=\num{0.0012}$ for architecture). (b) Versus covariance condition number $\mathrm{cond}\,\Sigma$ ($R^2=\num{0.959}$). Dotted line: $\share=1$.}
\label{fig:share_cloud}
\end{figure}

Aggregated by lattice size alone, the raw ensemble suggests a large architectural disparity: the median share of the single chain is \num{0.27}, \num{0.19}, \num{0.43}, and \num{0.83} at $N=6$, $8$, $10$, and $12$, against \num{0.88}, \num{0.92}, \num{0.93}, and \num{0.94} for the ladder, with the chain median rising at $N=10,12$ because the covariance check excludes its ill-conditioned amplifying tail. \Cref{fig:share_cloud} shows that this apparent gap disappears once configurations are organized by their end-to-end forward power gain $\Gfwd=10\log_{10}|S_{21}|^{2}$ (defined in~\cref{sec_realification}, with reverse isolation $\Iso=\Gfwd-\Grev$). Across all \num{15432} retained configurations, the architecture label alone explains only $R^2=\num{0.211}$ of the variance in $\share$, whereas $\Gfwd$ alone (\qty{2}{\dB} bins) explains $R^2=\num{0.978}$; adding the architecture label raises $R^2$ to \num{0.979}, an increment of only $\Delta R^2=\num{0.0012}$. Binning on $\log_{10}(\mathrm{cond}\,\Sigma)$ similarly explains $R^2=\num{0.959}$. Along this shared trend at $N=8$, the bin medians of $\share$ fall from \num{0.97} below \qty{0}{\dB} to \num{0.63} in \qtyrange{0}{20}{\dB}, \num{0.23} in \qtyrange{20}{40}{\dB}, \num{0.035} in \qtyrange{40}{60}{\dB}, and below \num{2e-3} above \qty{60}{\dB}. Below zero gain, the ladder-minus-chain median difference is $\num{-0.014}$ at $N=8$ (\num{260} chain, \num{2425} ladder points; \qty{95}{\percent} percentile bootstrap interval $\numrange{-0.021}{-0.0035}$) and $\num{-0.025}$ at $N=12$ (\num{163} and \num{2395} points; $\numrange{-0.034}{-0.016}$), both excluding zero under configuration resampling (Supplementary \cref{SI-sec:si_intervals}). At $N=8$ the median offset stays below \num{0.04} across every jointly populated gain band, while at $N=12$ it reaches \num{0.085} in \qtyrange{20}{40}{\dB} and \num{0.048} in \qtyrange{40}{60}{\dB} (Supplementary \cref{SI-tab:si_gain}).

Within the validated sample, the two architectures therefore differ primarily in the gain they extract from a given stability margin rather than in the share they retain at equal gain. From the loosest to the tightest margin band sampled, the single-chain median gain rises from \qty{+39}{\dB} to \qty{+96}{\dB} at $N=8$ and from \qty{-33}{\dB} to \qty{+97}{\dB} at $N=12$, driving $\share$ from \num{0.075} to below \num{1e-2} at $N=8$ and from \num{0.937} to \num{7.8e-3} at $N=12$. Over the same stability margins, the ladder medians stay between \qty{-58}{\dB} and \qty{+18}{\dB} and retain $\share=\numrange{0.38}{0.95}$ (Supplementary \cref{SI-tab:si_matched}); at the loosest margin for $N=12$, where both sit at negative gain, their shares nearly coincide (\num{0.937} versus \num{0.953}). Inter-chain coupling thus governs how much forward gain the stability constraint permits, shifting the operating point along the curve of~\cref{fig:share_cloud}(a) without altering the exchange rate between gain and mean-response share.

\subsection{The price of gain is paid in conditioning}

\begin{figure}[t]
\centering
\includegraphics[width=0.95\textwidth]{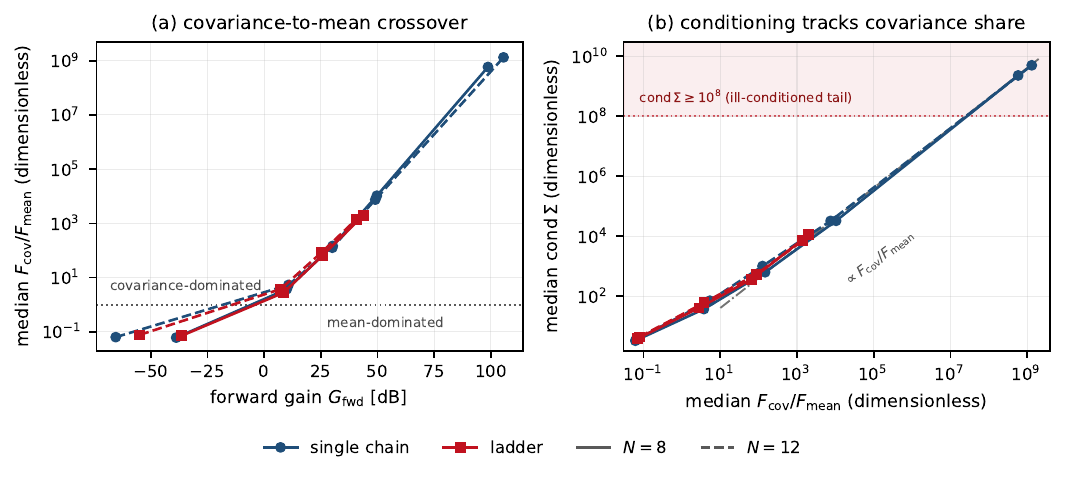}
\caption{(a) Median covariance-to-mean ratio $\Fcov/\Fmean$ of~\cref{eq:split} binned by forward gain, showing the crossover from mean-dominated to covariance-dominated Fisher information near \qty{0}{\dB} for the single chain (blue circles) and ladder (red squares) at $N=8$ (solid) and $N=12$ (dashed). (b) Median covariance condition number $\mathrm{cond}\,\Sigma$ plotted directly against $\Fcov/\Fmean$, demonstrating near-linear scaling $\mathrm{cond}\,\Sigma \propto \Fcov/\Fmean$ (dash-dotted slope-one guide) across ten orders of magnitude; shaded region marks the ill-conditioned tail ($\mathrm{cond}\,\Sigma \ge \num{1e8}$) reached only by the single chain.}
\label{fig:decomposition}
\end{figure}

As forward gain increases, the covariance term overtakes the mean response. At $N=8$, below \qty{0}{\dB}, $\Fcov/\Fmean$ is \num{0.06} for the chain and \num{0.07} for the ladder; across the next three \qty{20}{\dB} bands it climbs to \num{3.7}, \num{1.5e2}, and \num{1.1e4} for the chain and \num{2.8}, \num{65}, and \num{2.0e3} for the ladder. The covariance condition number rises in lockstep, from \num{3.2} to \num{37}, \num{620}, and \num{3.2e4} for the chain and from \num{3.9} to \num{40}, \num{360}, and \num{1.1e4} for the ladder (\cref{fig:decomposition}). Summarized over the full box, $\Fcov/\Fmean$ has median \num{37} ($N=6$) and \num{258} ($N=8$) for the chain versus \num{0.24} and \num{0.14} for the ladder.

This shared dependence on conditioning has a direct algebraic origin. Because both the drive $u=e_{N-1}$ and the signal generator $\dot A=-i e_{N-1}e_{N-1}^{\mathsf T}$ act exclusively on the last site of chain A, the mean-field derivative satisfies the exact identity $\dot\mu = -A^{-1}\dot A\mu = i(A^{-1})_{N-1,N-1}\mu$, independent of the chain length or inter-chain coupling. Substituting into~\cref{eq:split} gives
\begin{equation}
\frac{\Fmean}{\|\mu\|^2} \;=\; \left|(A^{-1})_{N-1,N-1}\right|^2 \,
\frac{\dot\mu_{\rm R}^{\mathsf T}\Sigma_{\rm R}^{-1}\dot\mu_{\rm R}}{\|\dot\mu_{\rm R}\|^2}.
\label{eq:rayleigh}
\end{equation}
The normalized mean response is therefore the product of the local Green function squared and the Rayleigh quotient $q$ of $\Sigma_{\rm R}^{-1}$ along the mean-field direction. This factorization separates two competing physical effects. At weak non-reciprocity and short length ($N=2$ in~\cref{tab:landscape}), modal hybridization lifts $|(A^{-1})_{N-1,N-1}|^2$ to \num{1.73} while the covariance remains nearly isotropic ($q=\num{0.588}$, $\mathrm{cond}\,\Sigma=\num{4.1}$), explaining the slight excess $\share=\num{1.024}>1$. At larger gain ($N=4,8$), however, the local susceptibility saturates near $\numrange{1.90}{2.00}$, whereas non-normal amplification concentrates both the steady-state field and the amplified noise fluctuations onto the same skin-localized mode; that mode becomes the largest-variance eigendirection of $\Sigma_{\rm R}$, depressing $q$ to \num{0.0089} ($N=4$) and \num{0.0295} ($N=8$) and pulling $\share$ down roughly as $1/\mathrm{cond}\,\Sigma$ irrespective of architecture. Consistent with this mechanism, the crossover $\Fcov=\Fmean$ in~\cref{fig:decomposition}(a) occurs precisely around unity forward gain: configurations with $\Fcov/\Fmean\in[\num{0.8},\num{1.25}]$ have median $\Gfwd=\qty{0.15}{\dB}$ for the ladder and \qty{1.21}{\dB} for the single chain. Below unity gain, upstream vacuum fluctuations are attenuated before reaching the read-out port and $\Fmean$ dominates; once directional gain exceeds unity, amplified upstream vacuum noise piles up at the read-out boundary alongside the signal, making $\Fcov$ dominant.

The same conditioning sets the numerical horizon of double-precision modeling. Every margin-stable ladder configuration agrees with the Bartels--Stewart solver to a median relative error of $\num{2.5e-11}$, whereas \qty{32}{\percent} of single-chain configurations fail the one-percent tolerance in the ill-conditioned amplifying tail ($\mathrm{cond}\,\Sigma\gtrsim\num{1e8}$; \cref{sec_reliability}). The apparent divergence of $\Fnorm$ toward the Hurwitz threshold therefore marks a loss of double-precision numerical control rather than a resolvable metrological divergence—a regime the opposing-skin ladder avoids because coupling caps its gain well below that tail.

\begin{figure}[t]
\centering
\includegraphics[width=\textwidth]{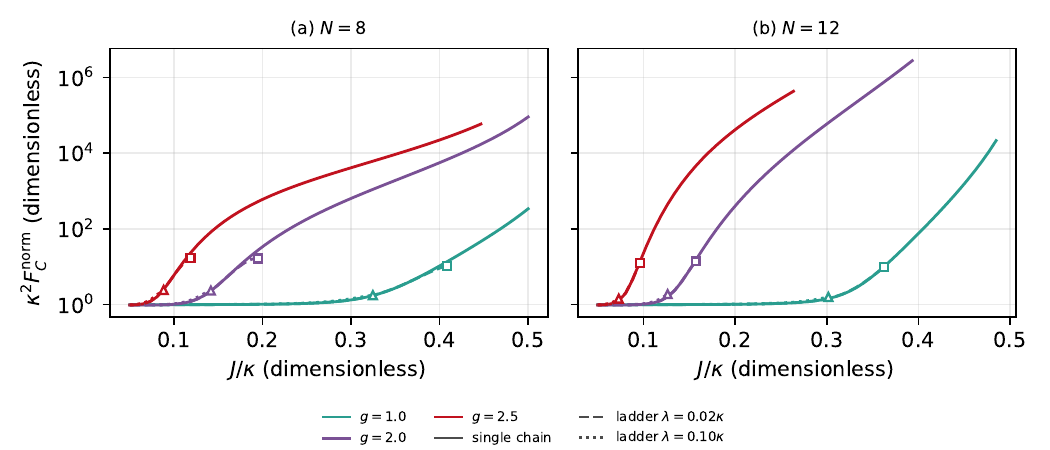}
\caption{Normalized Fisher information $\kappa^2\Fnorm$ along cuts in $J/\kappa$ at fixed non-reciprocity $g$, evaluated directly with the Bartels--Stewart Schur solver for points satisfying $\max\Real\nu<-0.05\kappa$, Lyapunov residual $\le\num{1e-12}$, and $\mathrm{cond}\,\Sigma\le\num{1e12}$. Solid: single chain; dashed (open square at termination): ladder at $\lambda=0.02\kappa$; dotted (open triangle at termination): ladder at $\lambda=0.10\kappa$; color encodes $g$. The ladder tracks the single chain closely and terminates early at its Hurwitz stability boundary, whereas the single chain grows until its covariance reaches the conditioning ceiling.}
\label{fig:cuts}
\end{figure}

Two further tests confirm this picture. Along one-dimensional cuts in $J$ at fixed non-reciprocity (\cref{fig:cuts}), computed directly with the Bartels--Stewart solver over points satisfying $\max\Real\nu<-0.05\kappa$, Lyapunov residual $\le\num{1e-12}$, and $\mathrm{cond}\,\Sigma\le\num{1e12}$, the ladder follows the single-chain curve closely before terminating at its Hurwitz limit—peaking between \num{9.65} and \num{16.9} at $\lambda=0.02\kappa$ and between \num{1.43} and \num{2.43} at $\lambda=0.10\kappa$. Across the same validated cuts, the single chain grows as an inverse power of the distance to threshold, with exponents \num{-2.42} (\qty{95}{\percent} interval $\numrange{-2.57}{-2.23}$, $R^2=\num{0.93}$) at $N=8, g=1$, \num{-4.96} ($\numrange{-5.85}{-4.38}$, $R^2=\num{0.86}$) at $N=8, g=2$, \num{-3.98} ($\numrange{-4.26}{-3.59}$, $R^2=\num{0.91}$) at $N=12, g=1$, and \num{-13.10} ($\numrange{-14.41}{-12.19}$, $R^2=\num{0.95}$) at $N=12, g=2$ before hitting the $\mathrm{cond}\,\Sigma=\num{1e12}$ horizon (Supplementary \cref{SI-tab:si_cuts}). Second, paired at identical $J$ and $g$ across jointly stable points for $N=6$--$12$ and $\lambda=\numrange{0.005}{0.1}\kappa$, the ladder and single chain agree in $\Fnorm$ to within one percent at the median and six percent at the ninetieth percentile (Supplementary \cref{SI-tab:si_ablation,SI-tab:si_ablation_b}). Adding the second chain neither creates nor destroys information at fixed local parameters; it shifts the stability boundary along~\cref{fig:share_cloud}.

\subsection{The operating points the ladder can reach}

Maximizing the total normalized information drives the ladder toward the stability boundary. \Cref{tab:landscape} reports the best of ten restarts of $1500$ Adam steps per size, each validated against the Bartels--Stewart solver. From $N=6$ upward every optimum lies within $0.05\kappa$ of the barrier ($\max\Real\nu/\kappa \in [\num{-0.094}, \num{-0.054}]$), whereas at $N=1$ and $N=2$ the spectrum remains deep inside the stable half-plane ($\num{-0.914}\kappa$ and $\num{-0.429}\kappa$). The normalized information spans $\num{0.995}$ at $N=1$ to $\num{95.6}$ at $N=14$ with no monotone trend in $N$, and for $N\ge4$ the optimizer pushes $\lambda$ close to its lower bound at four of seven sizes while reaching $\Gfwd$ up to \qty{+41.6}{\dB} and isolation $\Iso$ between \qty{71}{\dB} and \qty{89}{\dB}. Across $N\ge4$ the mean-response share drops to \numrange{0.018}{0.53}: maximizing total $F_C$ automatically trades away mean-field sensitivity for covariance sensitivity.

High gain is accompanied by strong eigenvector non-orthogonality, measured by the Petermann factor $K=1/|\langle\ell|r\rangle|^{2}$ of the biorthogonal drift eigenbasis. At the ladder optima for $N\ge4$, the minimum left--right overlap falls to \numrange{0.024}{0.088}, giving $K=\numrange{131}{1800}$, compared with $K=\numrange{5e6}{6e26}$ at the single-chain optima (Supplementary \cref{SI-sec:si_controls}). Across the entire sampled box, the ladder never exceeds \qty{51}{\dB} of forward gain (with \qty{92}{\dB} isolation), far below the single chain, which reaches \qty{+214}{\dB} in the validated envelope (Supplementary \cref{SI-tab:si_matched,SI-tab:si_envelope}). That ceiling stems from a closed-loop feedback instability specific to opposing skin orientations. An excitation amplified rightward along chain A ($\propto e^{+g(N-1)}$) crosses via $\lambda$ to chain B, is re-amplified leftward ($\propto e^{+g(N-1)}$), and returns to chain A, generating an effective round-trip loop gain $\propto\lambda^2 e^{2g(N-1)}$ that drives the ladder across the Hurwitz self-oscillation threshold. Equivalently, transforming to the similarity frame that symmetrizes each chain rescales the opposing-skin coupling as $\lambda e^{2gj}$, so the critical coupling for stability collapses roughly as $e^{-4g}$—from \num{0.214} at $g=1$ to \num{0.0006} at $g=2.5$ for $N=6$. Reversing the second chain so that both skins accumulate at the same boundary turns the return path into an attenuator ($\propto e^{-g(N-1)}$), removing the exponential loop gain and restoring single-chain growth to $\kappa^2\Fnorm=\num{5.7e8}$ at $N=12$ against \num{12.3} for the opposing ladder (Supplementary \cref{SI-sec:si_controls}). An opposing-skin ladder cannot be simultaneously strongly non-reciprocal and coupled.

\begin{table}[t]
\centering
\caption{Optima of the total resource-normalized information $\kappa^2\Fnorm$ at fixed size, from ten restarts of $1500$ Adam steps, each validated by the Bartels--Stewart solver check. $\phi$ is a gauge phase and is omitted; $\lambda$ is the inter-chain coupling; $\max\Real\nu/\kappa$ is the spectral abscissa of the drift; $\share$ and $\mathrm{cond}\,\Sigma$ are evaluated at each optimum. Nearest-neighbor hopping $(J,g)$ and end-to-end transmission $(\Gfwd,\Iso)$ are defined for $N\ge2$.}
\label{tab:landscape}
\footnotesize
\setlength{\tabcolsep}{2.5pt}
\begin{tabular}{@{}rccccccccc@{}}
\toprule
$N$ & $J/\kappa$ & $g$ & $\lambda/\kappa$ & $\kappa^2\Fnorm$ & $\share$ & $\mathrm{cond}\,\Sigma$ & $\max\Real\nu/\kappa$ & {$\Gfwd$ [\si{\dB}]} & {$\Iso$ [\si{\dB}]}\\
\midrule
1  & --           & --           & \num{0.0998} & \num{0.995} & \num{1.000} & \num{1.2}   & \num{-0.914} & --          & --          \\
2  & \num{0.474}  & \num{0.503}  & \num{0.0918} & \num{1.126} & \num{1.024} & \num{4.1}   & \num{-0.429} & \num{3.4}   & \num{8.5}   \\
4  & \num{0.407}  & \num{2.074}  & \num{0.0050} & \num{12.58} & \num{0.018} & \num{4.7e3} & \num{-0.156} & \num{40.2}  & \num{86.9}  \\
6  & \num{0.351}  & \num{1.729}  & \num{0.0056} & \num{38.71} & \num{0.040} & \num{5.9e3} & \num{-0.094} & \num{41.3}  & \num{86.7}  \\
8  & \num{0.293}  & \num{1.720}  & \num{0.0055} & \num{67.20} & \num{0.057} & \num{6.8e3} & \num{-0.070} & \num{41.6}  & \num{87.9}  \\
10 & \num{0.188}  & \num{1.992}  & \num{0.0107} & \num{40.80} & \num{0.125} & \num{1.8e3} & \num{-0.065} & \num{34.1}  & \num{77.8}  \\
12 & \num{0.439}  & \num{0.715}  & \num{0.0086} & \num{12.27} & \num{0.525} & \num{340}   & \num{-0.055} & \num{21.4}  & \num{71.1}  \\
14 & \num{0.173}  & \num{1.992}  & \num{0.0057} & \num{95.59} & \num{0.154} & \num{2.9e3} & \num{-0.063} & \num{34.9}  & \num{88.9}  \\
16 & \num{0.151}  & \num{2.012}  & \num{0.0147} & \num{23.92} & \num{0.303} & \num{560}   & \num{-0.054} & \num{24.3}  & \num{72.5}  \\
\bottomrule
\end{tabular}
\end{table}

\section{Discussion}

Two lessons follow, one methodological and one architectural, and neither is likely restricted to this specific lattice. Any linear read-out that extracts a large response from a fixed photon budget—whether non-reciprocal, parametric, or optomechanical—faces the same trade-off between gain and covariance conditioning along the signal mode. This makes~\cref{eq:rayleigh} and the gain-conditioned trend of~\cref{fig:share_cloud} a candidate design rule for linear amplifying sensors.

Because the classical Gaussian Fisher information splits into distinct mean-response and covariance-response terms, reporting $\Fmean$, $\Fcov/\Fmean$, and $\mathrm{cond}\,\Sigma$ separately is essential to see how much precision is supported by displacement measurements and where numerical conditioning limits precision. Comparing architectures via the mean-response share $\share$ also avoids the convention ambiguity of ranking total Fisher information across families whose Petermann factors differ by many orders of magnitude; here $K$ is reported as a complementary diagnostic of non-orthogonality rather than as a ranking weight. Equally important is realification bookkeeping: as~\cref{fig:metric_audit} demonstrates, an interleaved stride mismatch in a block-realified drift passes single-cavity tests silently while inverting the ranking of multi-mode arrays.

Within the validated parameter box, the empirical share-versus-gain relation ($R^2=\num{0.978}$, architecture increment $\Delta R^2=\num{0.0012}$) shows that forward gain governs the transfer of Fisher information from the mean field into the noise covariance. Inter-chain coupling provides an engineering knob to restrict the operating point to moderate gains and well-conditioned covariances under stability, rather than a mechanism that increases the mean-response share at matched gain.

These conclusions pertain to the classical Gaussian Fisher information of the linearized intracavity steady state under co-localized drive and sensing ports, and point to clear operational distinctions and future directions. Propagating the intracavity field through a specific input--output readout chain or evaluating the full Gaussian QFI requires an explicit detector model (Supplementary \cref{SI-sec:si_platform}). Even at the state level, adding isotropic excess noise $\Sigma\to\Sigma+n_{\mathrm{add}}I$ highlights the operational contrast between the two terms: one added quantum reduces the exact clean-cavity share from \num{1.000} to \num{0.500} (and the $N=1$ coupled ladder optimum from \num{1.000} to \num{0.504}), whereas at the amplifying $N=4$ single-chain optimum $\Fcov/\Fmean$ moves only from \num{4.6e5} at $n_{\mathrm{add}}=0$ to \num{4.1e5} at $n_{\mathrm{add}}=5$, leaving the amplifying regime strictly covariance-dominated. Extending the present framework beyond the sampled linear box ($N=6$--$12$, $J\in[0.05,0.5]\kappa$, $g\in[0.5,3]$, $\lambda\in[0.005,0.1]\kappa$) to closed synthetic-flux plaquettes that lift the gauge invariance of $\phi$, to spatially separated drive and sensing ports, or to quadratic and photon-counting read-outs capable of directly harvesting $\Fcov$ offers promising avenues for future work. Complementary metrological trade-offs in Floquet topological Wigner-negativity protection and ultrastrongly coupled cavity-loss limits are developed in companion studies targeting distinct physical platforms and observables \citep{Deuyekbe2026ResourceInvariant,Deuyekbe2026CavityLoss}.

For practical sensor design, the resulting rule is compact. Prefer operating regimes where the steady-state covariance remains well conditioned, since $\share$ and $\mathrm{cond}\,\Sigma$ move together and displacement measurements access only $\Fmean$. Never optimize total Fisher information without inspecting its mean--covariance split, because unconstrained maximization drives the system onto the stability boundary where little mean-response sensitivity remains. Finally, whenever a modified lattice is proposed to recover information lost to amplification, compare candidate architectures at matched forward gain: otherwise the comparison measures the gain rather than the architecture.

\section{Methods}

\subsection{Drift, steady state, scattering and Fisher information}
\label{sec_realification}

Using zero-based indexing $j=0,\dots,N-1$ per chain, chain A hops rightward ($j\to j+1$) with amplitude $J e^{+g}$ and leftward ($j+1\to j$) with $J e^{-g}$, while chain B carries the opposite signs. The two chains are linked site by site through a dissipative (anti-Hermitian in $H_{\rm eff}=iA$) coupling $\lambda e^{\pm i\phi}$. With uniform amplitude decay rate $\kappa$ (energy decay rate $2\kappa$) and frequency shift $\theta$ at the last site ($j=N-1$) of chain A, the complex $2N\times 2N$ drift matrix is
\begin{equation}
A \;=\;
\begin{pmatrix} A_{\rm A} & \lambda e^{i\phi}I_N\\ \lambda e^{-i\phi}I_N & A_{\rm B}\end{pmatrix},
\qquad
(A_{\rm A})_{jk} = -\kappa\,\delta_{jk} - i\theta\,\delta_{j,N-1}\delta_{k,N-1} + J e^{+g}\delta_{j,k+1} + J e^{-g}\delta_{j+1,k},
\label{eq:drift}
\end{equation}
with $A_{\rm B}$ carrying $g\to -g$, open boundary conditions, and drive $u=e_{N-1}$ on the last site of chain A. The steady-state mean and Hermitian covariance satisfy $\mu=-A^{-1}u$ and $A\Sigma+\Sigma A^{\dagger}=-2\kappa I$. For $N\ge2$, end-to-end transmission along chain A is evaluated in the input--output scattering convention $S_{21}=-\kappa_{\rm port}(A^{-1})_{N-1,0}$ and $S_{12}=-\kappa_{\rm port}(A^{-1})_{0,N-1}$ with symmetric external port coupling $\kappa_{\rm port}=\sqrt{2}\kappa\le 2\kappa$, yielding forward gain $\Gfwd=10\log_{10}|S_{21}|^2$, reverse gain $\Grev=10\log_{10}|S_{12}|^2$, and reverse isolation $\Iso=\Gfwd-\Grev$.

To evaluate the real Fisher functional~\cref{eq:split}, we realify the complex drift of dimension $m=2N$ into the $2m\times 2m$ block matrix
\begin{equation}
A_{\rm R} = \begin{pmatrix} \Real A & -\Imag A\\ \Imag A & \Real A\end{pmatrix},
\qquad
\mu_{\rm R} = \begin{pmatrix}\Real\mu\\ \Imag\mu\end{pmatrix},
\label{eq:realify}
\end{equation}
solving $A_{\rm R}\Sigma_{\rm R}+\Sigma_{\rm R}A_{\rm R}^{\mathsf T}=-2\kappa I_{2m}$ together with $\dot\mu_{\rm R}=-A_{\rm R}^{-1}\dot A_{\rm R}\mu_{\rm R}$ and the implicit Lyapunov equation for $\dot\Sigma_{\rm R}$. In~\cref{eq:realify}, site $j$ of chain A occupies real index $j$ and its imaginary partner occupies index $m+j$, so the drive sits at index $N-1$ and the detuning generator is $(\mathrm dA_{\rm R})_{N-1,\,m+N-1}=+1$, $(\mathrm dA_{\rm R})_{m+N-1,\,N-1}=-1$. Using an interleaved stride index $2(N-1)$ in place of $N-1$ (and $2(N-1)+1$ in place of $m+N-1$) produces the convention mismatch diagnosed in~\cref{fig:metric_audit}; the two coincide only for a single resonator ($m=N=1$). Every reported value uses~\cref{eq:realify}, and analytic derivatives agree with central finite differences in $\theta$ to below $10^{-6}$ relative error across all points of~\cref{tab:landscape}.

\subsection{Spectral Lyapunov solver and adjoint}

Solving the covariance Lyapunov equation dominates optimization cost, and the dense Kronecker form becomes prohibitive beyond $N=8$. Diagonalizing $A_{\rm R}=V\Lambda V^{-1}$ with eigenvalues $\nu_i$ gives
\begin{equation}
\Sigma_{\rm R} \;=\; V\Big[\frac{V^{-1}(-2\kappa I)V^{-\mathsf T}}{\nu_i+\nu_j}\Big]V^{\mathsf T},
\label{eq:spectral}
\end{equation}
where the Hurwitz stability margin guarantees $|\nu_i+\nu_j|\ge2\times\num{0.05}\kappa$. The forward cost drops from \qty{3816}{\milli\second} to \qty{7.3}{\milli\second} per evaluation at $N=8$. Because automatic differentiation does not differentiate through general eigenvectors, \cref{eq:spectral} is wrapped in a custom vector--Jacobian product whose backward pass solves the transposed Lyapunov equation; its derivation, Kronecker cross-check, and timing are given in Supplementary \cref{SI-sec:si_solver}. All arithmetic runs in double precision on CPU.

\subsection{Numerical reliability of the covariance}
\label{sec_reliability}

Although~\cref{eq:spectral} is exact in exact arithmetic and differentiable, its forward error grows with the eigenvector condition number of a strongly non-normal drift. Every configuration reported here is therefore recomputed with the independent Bartels--Stewart Schur solver \citep{Bartels1972}, which avoids eigenvector inversion altogether, and retained only when the two routes agree within one percent in $\Fnorm$. Across the \num{34560} grid points, \num{17599} satisfy $\max\Real\nu<0$ (\num{16369} passing the solver check) and \num{16646} satisfy the $0.05\kappa$ stability margin (\num{3780} single-chain, \num{12866} ladder), of which \num{15432} (\num{2566} single-chain, \num{12866} ladder) pass the Bartels--Stewart check and enter~\cref{fig:share_cloud,fig:decomposition} and Supplementary Tables S4--S8. Every margin-stable ladder configuration passes with a median relative difference of $\num{2.5e-11}$, whereas \qty{32}{\percent} of single-chain points fail, concentrated in the ill-conditioned amplifying tail ($\mathrm{cond}\,\Sigma\gtrsim\num{1e8}$, although \num{462} retained single-chain points above \num{1e8} still pass the relative check). Failed points are excluded from all reported statistics and conclusions.

Two additional checks bound residual error. The single-chain spectrum satisfies $\Real\nu=-\kappa+2J\cos[\pi/(N+1)]$ independently of $g$, reproduced to $\num{2.8e-3}$ across the scan, confirming that distance to the Hurwitz barrier is set by $J$ alone while non-reciprocity $g$ sets the gain obtained at that distance. Furthermore, fifty-digit arbitrary-precision Lyapunov checks at extreme configurations yield residuals below $\num{1e-8}$ at every passing point and orders of magnitude larger at failing points.

\subsection{Optimal control and statistics}

Optimization maximizes $\kappa^2\Fnorm$ with Adam \citep{Kingma2014} on a sigmoid-mapped box, $J\in[0.05,0.5]\kappa$, $g\in[0.5,3]$, $\lambda\in[0.005,0.1]\kappa$, $\phi\in[0,2\pi)$, with learning rate \num{0.003} and a one-sided barrier penalty $a\max(\Real\nu_{\max}/\kappa+\num{0.05},0)$ ($a=10^8$). Gradients are computed by reverse-mode automatic differentiation \citep{Bradbury2018,Leung2017}. Each entry in~\cref{tab:landscape} is the best of ten restarts of $1500$ Adam steps from widened initializations (base seed \num{20260823}, stride \num{7919}), reported as the best solution found under this protocol. The grid comparison of~\cref{fig:share_cloud} and Supplementary \cref{SI-tab:si_matched} evaluates \num{32} values of $J$, \num{30} of $g$, and \num{8} of $\lambda$ per size using central differences in $\theta$ with step \num{1e-5}, filtered by $\max\Real\nu<-0.05\kappa$ and the one-percent Bartels--Stewart check of~\cref{sec_reliability}. The clean-cavity benchmark~\cref{eq:cavity} is exact.

\section*{Data availability}

The data supporting the findings of this study are available from the corresponding author on reasonable request and will be released publicly, with a registered identifier, upon acceptance.

\section*{Code availability}

The analysis scripts and environment specification are available from the corresponding author on reasonable request and will be released publicly with the data, with a registered identifier, upon acceptance.

\section*{Acknowledgments}

This research received no specific grant from any funding agency in the public, commercial, or not-for-profit sectors. Computational resources were provided by the penavoraserver cluster (ThinkStation P720, \qty{128}{\giga\byte} RAM, \num{48} cores) through Penabei Samafou.

\emph{Use of AI.} A large language model assisted with code development and with verifying consistency between the manuscript and numerical outputs. The authors are responsible for the scientific content and the final submitted text.

\section*{Author contributions}

J.D. performed the numerical work and the optimal-control implementation. J.D. and P.D. designed the study and analyzed the results. N.M. contributed the scattering analysis and the platform parameters. S.G.N.E. supervised the project, wrote the manuscript with J.D., and verified the numerical analysis and the bibliography. All authors discussed the results and commented on the manuscript.

\section*{Competing interests}

The authors declare no competing interests.

\bibliographystyle{unsrtnat}
\bibliography{04_track_C_nhse_references}

\end{document}


\title{Supplementary Information for ``Forward gain tracks the mean-response share of the Fisher information in non-Hermitian bosonic ladders''}

\author{Juste Deuyekbe}
\affiliation{Department of Physics, Faculty of Science, University of Ngaound\'er\'e, P.O. Box 454, Ngaound\'er\'e, Cameroon}

\author{P. Djorw\'e}
\email{djorwepp@gmail.com}
\affiliation{Department of Physics, Faculty of Science, University of Ngaound\'er\'e, P.O. Box 454, Ngaound\'er\'e, Cameroon}
\affiliation{Stellenbosch Institute for Advanced Study (STIAS), Wallenberg Research Centre at Stellenbosch University, Stellenbosch 7600, South Africa}

\author{Nsangou Mama}
\email{mnsangou@yahoo.com}
\affiliation{Department of Physics, Faculty of Science, University of Maroua, P.O. Box 46, Maroua, Cameroon}

\author{S. G. Nana Engo}
\email{serge.nana-engo@facsciences-uy1.cm}
\affiliation{Department of Physics, Faculty of Science, University of Yaound\'e~I, P.O. Box 812, Yaound\'e, Cameroon}

\maketitle

This Supplement fixes the conventions used in the main text, documents the
realification indexing benchmark of Sec.~2.2 of the main text, gives the solver
and the optimization protocol, and collects the tables behind
Figs.~2--5 of the main text: the barrier-matched comparison, the gain-binned
trend, the attainable envelope, the fixed-parameter ablation, the threshold
cuts and the control ablations. The last section states what the model does and
does not represent.

\section{Model and conventions}
\label{sec:si_model}

Each chain has $N$ sites with annihilation operators $a_j$, $j=0,\dots,N-1$,
nearest-neighbor hopping $J e^{\pm g}$, and photon loss $\kappa$ at every site~\cite{Hatano1997,Yao2018}.
The two chains carry opposite signs of the non-reciprocity $g$, so their skin
accumulations point at opposite boundaries, and they are coupled site by site
with amplitude $\lambda$ and phase $\phi$~\cite{McDonald2020Exponential,Metelmann2015}. In the frame rotating at the drive
frequency the linearized drift of the ladder, at fixed signal detuning $\theta$
applied to the last site of the first chain, is the $2N\times2N$ matrix
\begin{equation}
A \;=\;
\begin{pmatrix}
A_{\rm A} & \lambda e^{i\phi}\mathbb{1}_N\\
\lambda e^{-i\phi}\mathbb{1}_N & A_{\rm B}
\end{pmatrix},
\qquad
\left(A_{\rm A}\right)_{jk} = -\kappa\,\delta_{jk}
+ J e^{g}\,\delta_{j,k+1} + J e^{-g}\,\delta_{j+1,k}
- i\theta\,\delta_{j,N-1}\delta_{k,N-1},
\label{eq:si_drift}
\end{equation}
with the opposite hopping asymmetry ($g\to -g$) in $A_{\rm B}$. A single chain is the same
object with $\lambda=0$ and one block. The drive enters at the last site of the first
chain, $u_{N-1}=1$, and the diffusion matrix is $2\kappa\mathbb{1}_{2N}$, the vacuum
contribution of the two chains. The steady state is Gaussian, with mean
\begin{equation}
\mu \;=\; -A^{-1}u,
\label{eq:si_mean}
\end{equation}
and Hermitian covariance solving the Lyapunov equation $A\Sigma+\Sigma A^{\dagger}=-2\kappa
\mathbb{1}$.

Because $A$ is complex, the numerical problem is posed in real form. Writing
$z = x\otimes 1 + y\otimes i$ and stacking the real and imaginary parts of the
whole vector, the realified drift of a matrix of dimension $m=2N$ is
\begin{equation}
A_{\rm R} \;=\;
\begin{pmatrix}
\Real A & -\Imag A\\
\Imag A & \Real A
\end{pmatrix}
\in\mathbb{R}^{2m\times2m},
\label{eq:si_realify}
\end{equation}
the realified drive is $u_{\rm R}=(u,0)$, and the realified covariance solves
$A_{\rm R}\Sigma_{\rm R}+\Sigma_{\rm R}A_{\rm R}^{\mathsf T}
=-2\kappa\mathbb{1}_{2m}$. In this layout the real part of mode $j$ occupies
index $j$ and its imaginary part occupies index $m+j$: \emph{the block offset is
$m$, not the site count $N$}. This is the point of Sec.~2.2 of the main text and
of Sec.~\ref{sec:si_index} below.

The Fisher information of the Gaussian steady state about $\theta$ is the sum of
a mean-response and a noise-response term~\cite{Pinel2013,Safranek2019},
\begin{equation}
F_C \;=\; \dot\mu^{\mathsf T}\Sigma^{-1}\dot\mu \;+\;
\tfrac12\Tr\!\left[\left(\Sigma^{-1}\dot\Sigma\right)^{2}\right]
\;\equiv\; \Fmean+\Fcov ,
\label{eq:si_split}
\end{equation}
where the dot is $\mathrm d/\mathrm d\theta$ and we drop the subscript ${\rm R}$ on realified vectors and matrices below whenever no ambiguity arises. Both derivatives are evaluated
analytically: $\dot\mu$ solves $A\,\dot\mu=-\dot A\mu$ and $\dot\Sigma$ solves
$A\,\dot\Sigma+\dot\Sigma A^{\mathsf T}=-(\dot A\Sigma+\Sigma\dot A^{\mathsf T})$,
with $\dot A$ the rotation generator of the detuning at the signal site.
Equation~\eqref{eq:si_split} is the Gaussian-state Fisher information of the
intracavity steady state about $\theta$, in units where the vacuum noise is one:
the classical information of a Gaussian outcome distribution with mean $\mu$ and
covariance $\Sigma$. It is not the information of a photon-counting record, and
the field is not propagated to the output port here; the port enters only through
the dilution factor of \cref{sec:si_platform}, and the phase-insensitive noise of a
real detection chain adds to the covariance term in quadrature there.

Resource normalization divides \eqref{eq:si_split} by the steady-state occupation
$\|\mu\|^2=\sum_j|\mu_j|^2$, counted on both chains for the ladder and on the
single chain for the baseline, and defines $\Fnorm=F_C/\|\mu\|^2$. The reference
configuration is one cavity driven and read at the same port, $A=-\kappa-i\theta$.
Its covariance is $\Sigma=\mathbb{1}$ \emph{independently of $\theta$} (the diffusion matrix is $2\kappa\mathbb{1}$ and the drift is $-\kappa$, so the normalization is the one in which a cold cavity has unit covariance), so
$\Fcov\equiv0$, the covariance is perfectly conditioned, and
\begin{equation}
\kappa^2\Fnorm \;=\; \frac{1}{1+(\theta/\kappa)^2},
\label{eq:si_ceiling}
\end{equation}
exactly, at every drive amplitude. At the operating point $\theta_0=0.1\kappa$
used throughout this work, $\kappa^2\Fnorm=\num{0.9901}$. The mean-response share
\begin{equation}
\share \;=\; \frac{\kappa^2\Fmean/\|\mu\|^2}{1/[1+(\theta_0/\kappa)^2]}
\label{eq:si_share}
\end{equation}
therefore measures the fraction of the clean-cavity response that an
architecture retains in its mean field. It equals one when the covariance does not depend on the parameter and the normalized mean response reaches the cavity value, and tends to zero for an architecture whose information is carried by the parameter dependence of its noise. Isotropy alone is not sufficient: a lattice with an isotropic covariance $\Sigma=c(\theta)\mathbb{1}$ whose scale $c(\theta_0)>1$ depends on $\theta$ carries both $\Fcov\neq0$ and a reduced mean-response share $\share<1$.

\subsection{Notation}

\begin{table}[htbp]
\caption{Parameters of the model, their meaning and their values at the
operating points analyzed here. Frequencies are in units of $\kappa$; the
dimensional values in the last column assume $\kappa/2\pi=\qty{1.7}{MHz}$.}
\label{tab:si_glossary}
{\footnotesize
\begin{ruledtabular}
\begin{tabular}{@{}llll@{}}
Symbol & Meaning & Values used & Dimensional \\
\hline
$N$ & sites per chain & $1$--$16$ & -- \\
$J$ & reciprocal hopping & $0.05$--$0.50\,\kappa$ & \qtyrange{85}{850}{kHz} \\
$g$ & non-reciprocity & $0.50$--$3.00$ & $\gNR=2J\sinh g$ \\
$\lambda$ & inter-chain coupling & $0.005$--$0.100\,\kappa$ & \qtyrange{8}{170}{kHz} \\
$\phi$ & inter-chain phase & gauge & -- \\
$\kappa$ & photon loss per site & $1$ & \qty{1.7}{MHz} \\
$\kappa_{\rm port}$ & symmetric port coupling & $\sqrt{2}\,\kappa\le 2\kappa$ & $\sqrt{2}\kappa/2\pi$ \\
$\theta_0$ & signal detuning & $0.1\,\kappa$ & \qty{170}{kHz} \\
$\nu_i$, $\Real\nu_{\max}$ & drift eigenvalues, abscissa & $\Real\nu_{\max} < -0.05\,\kappa$ & \si{MHz} \\
$\Fmean$, $\Fcov$ & mean, covariance Fisher terms & -- & $\kappa^{-2}$ \\
$\Fnorm$ & normalized Fisher information & -- & $\kappa^{-2}$ per photon \\
$\share$ & mean-response share & unitless, near $1$ at low gain & -- \\
$q$ & Rayleigh quotient of $\Sigma_{\rm R}^{-1}$ & $\dot\mu_{\rm R}^{\mathsf T}\Sigma_{\rm R}^{-1}\dot\mu_{\rm R}/\|\dot\mu_{\rm R}\|^2$ & -- \\
$K$ & Petermann factor & $1/|\langle\ell_i|r_i\rangle|^2\ge 1$ & -- \\
$\Gfwd$, $\Grev$ & forward, reverse port gain & \si{\dB} & \si{\dB} \\
$\Iso$ & reverse isolation & \si{\dB} & \si{\dB} \\
\end{tabular}
\end{ruledtabular}}
\end{table}

Because $g$ enters only through $e^{\pm g}$, $\phi$ is a gauge label of the
ladder: it is removed by $a_j\to e^{-i\phi}a_j$ on the second chain. The
physically meaningful control parameters are therefore $(J,g,\lambda)$ together
with the chain length $N$, and the optimizer is free in $\phi$ without
consequence, which is why the recovered values of $\phi$ differ between
independent runs that reach the same information.

\section{The read-out index of the realified state}
\label{sec:si_index}

Sec.~2.2 of the main text diagnoses a realification indexing mismatch in which the
drive and detuning derivative are placed at the interleaved index $2(N-1)$ instead
of the block-aligned index $N-1$. In the layout of \eqref{eq:si_realify}, mode $j$
of chain A occupies index $j$ and its imaginary partner sits at $m+j$ ($m=2N$), so
the interleaved stride $(2j,2j+1)$ coincides with the block convention only for a
single resonator ($m=N=1$). Table~\ref{tab:si_audit} compares the two evaluations
across ten representative ladder operating points: for the amplifying benchmark
rows (500-step starts) the stride mismatch changes $\kappa^2\Fnorm$ by factors of
up to $170$ and reverses the ordering of amplifying and deep-stability points; for
the deep-stability rows (1500-step continuations) the difference is at the percent
level because those configurations lie near the clean-cavity benchmark
\eqref{eq:si_ceiling} where the mean response dominates.

The ten benchmark points span two optimization budgets, labeled accordingly in
Table~\ref{tab:si_audit}. The six short-budget rows are ten-restart $500$-step
Adam optima; the four converged rows continue the same starts to $1500$ steps at
identical learning rate, barrier, and seed base. Under the interleaved index
$2(N-1)$, extending the budget moves $\kappa^2\Fnorm$ at $N=8$, $12$, and $16$
from \num{0.10}, \num{0.050}, and \num{0.043} to \num{0.970}, \num{0.968}, and
\num{0.969}, whereas under the block-aligned index $N-1$ the same rows read
\num{2.44}, \num{5.77}, and \num{7.25} at $500$ steps and \num{1.00}, \num{1.00},
and \num{1.00} after continuation. Gradient norms at the $500$-step points are
\numrange{0.43}{0.60} against \numrange{0.016}{0.104} in the deep-stability rows.
Throughout this article ``optimum'' denotes the best restart found under the
stated protocol rather than a certified stationary point.

\begin{table}[htbp]
\caption{Benchmark of realification indexing on $\kappa^2\Fnorm$ at ten
representative ladder operating points, comparing the block-aligned read-out
index $N-1$ against the interleaved index $2(N-1)$, together with their ratio
(block over interleaved).}
\label{tab:si_audit}
\begin{ruledtabular}
\footnotesize
\begin{tabular}{@{}rccccccc@{}}
$N$ & runs & $J/\kappa$ & $g$ & $\lambda/\kappa$ &
$\kappa^2\Fnorm$ at index $N-1$ & $\kappa^2\Fnorm$ at index $2(N-1)$ & ratio \\
\hline
6 & 500-step start & \num{0.105} & \num{1.250} & \num{0.016} & \num{0.989} & \num{0.937} & \num{1.056}\\
8 & 500-step start & \num{0.354} & \num{0.973} & \num{0.070} & \num{2.436} & \num{0.101} & \num{24.051}\\
10 & 500-step start & \num{0.078} & \num{1.001} & \num{0.079} & \num{0.995} & \num{0.983} & \num{1.012}\\
12 & 500-step start & \num{0.345} & \num{1.024} & \num{0.028} & \num{5.768} & \num{0.050} & \num{114.545}\\
14 & 500-step start & \num{0.087} & \num{0.797} & \num{0.077} & \num{0.997} & \num{0.987} & \num{1.010}\\
16 & 500-step start & \num{0.287} & \num{1.226} & \num{0.021} & \num{7.252} & \num{0.043} & \num{170.299}\\
6 & 1500-step cont. & \num{0.062} & \num{0.700} & \num{0.096} & \num{0.997} & \num{0.998} & \num{0.999}\\
8 & 1500-step cont. & \num{0.142} & \num{0.640} & \num{0.045} & \num{1.004} & \num{0.969} & \num{1.036}\\
12 & 1500-step cont. & \num{0.142} & \num{0.664} & \num{0.018} & \num{1.003} & \num{0.968} & \num{1.037}\\
16 & 1500-step cont. & \num{0.127} & \num{0.762} & \num{0.015} & \num{1.000} & \num{0.969} & \num{1.032}\\

\end{tabular}
\end{ruledtabular}
\end{table}

Table~\ref{tab:si_audit} demonstrates three artifacts caused by the interleaved
stride mismatch: (i) an apparent upper bound $\kappa^2\Fnorm\le1$, whereas
block-aligned evaluation yields $\num{2.44}$, $\num{5.77}$, and $\num{7.25}$ at
stable, feasible points; (ii) an inverted two-band landscape in which amplifying
points appear suppressed to $10^{-2}$ instead of exceeding the deep-stability
points by factors of two to seven; and (iii) an apparent ``single attractor''
near $\num{0.97}$, which is simply the clean-cavity benchmark
\eqref{eq:si_ceiling} at $\theta_0=0.1\kappa$ approached whenever optimization
stalls near the decoupled limit.

\section{Spectral Lyapunov solver and its adjoint}
\label{sec:si_solver}

Every evaluation of \eqref{eq:si_split} needs two Lyapunov solves, and the
optimization needs their derivative with respect to the four control
parameters. Solving $A_{\rm R}\Sigma_{\rm R}+\Sigma_{\rm R} A_{\rm R}^{\mathsf T}=R$ through the $m^2\times m^2$
Kronecker operator costs $\mathcal{O}(m^6)$ and dominated the runtime: at $N=8$ one
forward evaluation took \qty{3816}{ms} and one gradient step \qty{2976}{ms}. We
instead factorize the real drift once, $A_{\rm R}=V\Lambda V^{-1}$ with eigenvalues $\nu_i$, and use
\begin{equation}
\Sigma_{\rm R} \;=\; V\left[\frac{V^{-1}RV^{-\mathsf T}}{\nu_i+\nu_j}\right]V^{\mathsf T},
\label{eq:si_spectral}
\end{equation}
which is exact whenever no pair of eigenvalues of $A_{\rm R}$ sums to zero. The Hurwitz
constraint enforces $\Real\nu_i<-0.05\kappa$ at every accepted point,
so $\nu_i+\nu_j$ is bounded away from zero by $0.1\kappa$. That bound
governs the denominators only. The forward error of \eqref{eq:si_spectral} is set
by the conditioning of $V$ and of the resulting covariance, which both grow with
the non-normality of the drift and are not controlled by the barrier; for the
amplifying configurations of the single chain the covariance conditioning reaches
$10^{8}$ and beyond, and the route then loses digits. The solver is used because
it is differentiable and fast, not because it is uniformly accurate, and every quoted configuration is checked against an independent Schur-based solve as described in~\cref{sec:si_reliability}. The spectral route reduces a forward evaluation to \qty{7.3}{ms} and a gradient step to \qty{1.2}{ms}, a factor
of about $500$ and $2500$ respectively, and makes the campaign of
Table~\ref{tab:si_landscape} a matter of minutes.

Automatic differentiation cannot be applied to \eqref{eq:si_spectral} directly,
because the derivative of an eigenvector is not implemented in the autodiff
backend we use~\cite{Bradbury2018,Leung2017}. We therefore supply it. With $\Sigma_{\rm R}$ and the adjoint variable
$Y=\partial\ell/\partial\Sigma_{\rm R}$ held fixed, the variation of
\eqref{eq:si_spectral} is
$A_{\rm R}\,\delta\Sigma_{\rm R}+\delta\Sigma_{\rm R}\,A_{\rm R}^{\mathsf T}=-(\delta A_{\rm R}\,\Sigma_{\rm R}+\Sigma_{\rm R}\,\delta A_{\rm R}^{\mathsf T})$,
so the pullback with respect to $A_{\rm R}$ is
\begin{equation}
\bar A_{\rm R} \;=\; Y\Sigma_{\rm R}^{\mathsf T}+Y^{\mathsf T}\Sigma_{\rm R} ,
\label{eq:si_adjoint}
\end{equation}
obtained by solving the transposed Lyapunov problem for $Y$ with the same
factorization. Equation~\eqref{eq:si_adjoint} reproduces the dense Kronecker gradient to machine precision (relative accuracy $2.3\times10^{-14}$).

\section{Numerical reliability of the covariance}
\label{sec:si_reliability}

The realified covariance solves $A_{\rm R}\Sigma_{\rm R}+\Sigma_{\rm R} A_{\rm R}^{\mathsf T}=-2\kappa\mathbb{1}$, and the
optimizer evaluates it via the spectral formula~\eqref{eq:si_spectral}. The formula is exact in exact arithmetic, but its forward error is governed
by the condition number of the eigenvector matrix rather than by the spectral
gap: for a strongly non-normal drift the eigenvalues separate much faster than
the eigenvectors do, so a configuration can sit comfortably inside the barrier and
still return a covariance that is wrong by orders of magnitude.

Every configuration quoted in this work is therefore recomputed by an independent
route, the Bartels--Stewart algorithm~\cite{Bartels1972}, which reduces $A_{\rm R}$ by Schur vectors and
solves the resulting triangular equation without touching the eigenvectors. A
point is retained only when it satisfies the stability margin $\max\Real\nu<-0.05\kappa$ and the two routes agree to within one percent in $\Fnorm$. Out of the \num{34560} grid evaluations, \num{17599} satisfy $\max\Real\nu<0$ (\num{3840} single-chain, \num{13759} ladder, of which \num{16369} pass the solver check) and \num{16646} satisfy the stricter $0.05\kappa$ margin (\num{3780} single-chain, \num{12866} ladder). Of those \num{16646} margin-stable configurations, \num{15432} pass the Bartels--Stewart check and enter the analysis: all \num{12866} ladder configurations pass with a median relative difference of $\num{2.5e-11}$, while \num{2566} single-chain configurations pass (\qty{32}{\percent} fail overall). Failures are concentrated in the ill-conditioned amplifying tail ($\mathrm{cond}\,\Sigma\gtrsim\num{1e8}$, although \num{462} retained single-chain configurations above $\num{1e8}$ still pass the relative check). Failing points are excluded from all statistics and plots, and the divergence toward threshold is accordingly described as a loss of double-precision numerical control rather than a resolvable metrological prediction.

Three further checks bound the residual risk. First, the spectrum of a single
chain in this geometry is $\Real\nu=-\kappa+2J\cos[\pi/(N+1)]$ for every
$g$, so the distance to the barrier is the hopping alone; the measured values
reproduce the closed form to $\num{2.8e-3}$ over the whole scan, which certifies
that the drift itself, and not only the covariance, is being assembled
correctly. Second, at the configurations that pass the one percent test the
residual of the Lyapunov equation, evaluated in fifty-digit arithmetic, is below
$\num{1e-8}$; at those that fail it exceeds that figure by orders of magnitude.
Third, the analytic and finite-difference derivatives of the covariance, which
were used as an internal check of the adjoint of
\cref{sec:si_solver}, share the spectral solver and therefore validate the
derivative only, not the value; the Bartels--Stewart comparison above is the test
of the value.

\section{Optimization protocol and the landscape}
\label{sec:si_landscape}

The control vector is $x=(J,g,\lambda,\phi)$, mapped into the box
$J\in[0.05,0.5]\kappa$, $g\in[0.5,3]$, $\lambda\in[0.005,0.1]\kappa$,
$\phi\in[0,2\pi)$ by a sigmoid, and optimized by Adam~\cite{Kingma2014} at a learning rate of
$0.003$ with the loss
\begin{equation}
\ell(x) \;=\; -\kappa^2\Fnorm(x)\;+\;10^{8}\max\!\left(\Real\nu_{\max}(x)/\kappa+\num{0.05},\,0\right),
\label{eq:si_loss}
\end{equation}
which is finite, vanishes inside the stability margin, and grows in proportion to
the violation outside it; it is not differentiable where the violation vanishes.
Each size is optimized from ten independent random starts, with $1500$ steps each
and a common seed base, and the best restart is reported as the best solution found
under this protocol. Table~\ref{tab:si_landscape}
gives the outcome. From $N=6$ upward every recovered optimum sits within
$0.044\kappa$ of the barrier and at the four largest sizes within
$0.015\kappa$; the two smallest sizes stay far from it, at $-0.914\kappa$ and
$-0.429\kappa$, because with one or two modes the hopping alone moves the
spectrum and the hopping is bounded below. The non-reciprocity is not pinned at
the edge of its box anywhere: it spans $0.50$ to $2.07$ over $N=2$--$16$,
with $N=12$ at $0.715$. In the sampled optima, no single control parameter is consistently pinned to a
box boundary, and the operating point it finds is a compromise between
gain, coupling and stability rather than a corner of the box.

\begin{table}[htbp]
\caption{Best restart at each size: control parameters, resource-normalized
Fisher information, mean-response share, covariance-to-mean ratio, Petermann
factor, largest real part of the drift spectrum $\Real\nu_{\max}/\kappa$, forward gain, reverse isolation
and wall-clock time for the ten restarts. Frequencies in units of $\kappa$; gains
in \si{\dB}. Nearest-neighbor hopping $(J,g)$ and end-to-end transmission $(\Gfwd,\Iso)$ are defined for $N\ge2$ and left blank at $N=1$.}
\label{tab:si_landscape}
{\scriptsize\setlength{\tabcolsep}{2.5pt}%
\begin{ruledtabular}
\begin{tabular}{@{}rccccccccccc@{}}
$N$ & $J/\kappa$ & $g$ & $\lambda/\kappa$ & $\kappa^2\Fnorm$ & $\share$ &
$\Fcov/\Fmean$ & $K$ & $\Real\nu_{\max}/\kappa$ & $\Gfwd$ & $\Iso$ & $t$ (s) \\
\hline
1 & -- & -- & \num{0.100} & \num{0.995} & \num{1.000} & \num{5e-3} & \num{1.000} & \num{-0.914} & -- & -- & \num{7.200}\\
2 & \num{0.474} & \num{0.503} & \num{0.092} & \num{1.126} & \num{1.024} & \num{0.110} & \num{1.458} & \num{-0.429} & \num{3.382} & \num{8.512} & \num{7.900}\\
4 & \num{0.407} & \num{2.074} & \num{5.009e-3} & \num{12.577} & \num{0.018} & \num{708.478} & \num{1801.163} & \num{-0.156} & \num{40.232} & \num{86.941} & \num{13.200}\\
6 & \num{0.351} & \num{1.729} & \num{5.619e-3} & \num{38.710} & \num{0.040} & \num{977.415} & \num{892.214} & \num{-0.094} & \num{41.274} & \num{86.662} & \num{19.800}\\
8 & \num{0.293} & \num{1.720} & \num{5.530e-3} & \num{67.198} & \num{0.057} & \num{1196.975} & \num{757.737} & \num{-0.070} & \num{41.604} & \num{87.938} & \num{29.700}\\
10 & \num{0.188} & \num{1.992} & \num{0.011} & \num{40.805} & \num{0.125} & \num{329.498} & \num{231.151} & \num{-0.065} & \num{34.066} & \num{77.800} & \num{45.700}\\
12 & \num{0.439} & \num{0.715} & \num{8.573e-3} & \num{12.273} & \num{0.525} & \num{22.611} & \num{134.602} & \num{-0.055} & \num{21.398} & \num{71.147} & \num{65.200}\\
14 & \num{0.173} & \num{1.991} & \num{5.689e-3} & \num{95.587} & \num{0.154} & \num{626.664} & \num{639.760} & \num{-0.063} & \num{34.877} & \num{88.921} & \num{93.500}\\
16 & \num{0.151} & \num{2.012} & \num{0.015} & \num{23.922} & \num{0.303} & \num{78.784} & \num{130.574} & \num{-0.054} & \num{24.321} & \num{72.550} & \num{126.000}\\

\end{tabular}
\end{ruledtabular}}
\end{table}

The landscape has a remark that the main text uses. The optimum is not a
saturation point: the information keeps increasing as the barrier is
approached, and the value of $\Fnorm$ recovered at a given size is a statement
about how close the optimizer is allowed to sit, not about the architecture.
The chain lengths from $N=1$ to $N=16$ therefore differ by two orders of
magnitude in $\kappa^2\Fnorm$ without any monotone trend, and the ladder optimum at
$N=1$, $\kappa^2\Fnorm=\num{0.995}$, reproduces the single-cavity value
\eqref{eq:si_ceiling} to within half a percent, which is what one site and a
residual coupling of $0.1\kappa$ should give.

\section{Barrier-matched comparison}
\label{sec:si_matched}

Comparing two architectures at their respective optima compares two different
distances to the stability threshold, which is the dominant variable in the
problem. We therefore bin the sampled box by the distance to the barrier,
$|\Real\nu_{\max}|/\kappa$, and compare the two families band by band. The
box is the same for both: $J=0.05$--$0.5\kappa$ (32 values), $g=0.5$--$3.0$ (30
values), and for the ladder eight values of $\lambda=0.005$--$0.1\kappa$, which evaluates
\num{34560} parameter points. Of these, \num{17599} satisfy $\max\Real\nu<0$ (\num{3840} single-chain, \num{13759} ladder) and \num{16646} satisfy the stability margin $\Real\nu_{\max}<-0.05\kappa$ (\num{3780} single-chain, \num{12866} ladder). Combining this margin with finite positive information and the one-percent Bartels--Stewart solver check of \cref{sec:si_reliability} retains \num{15432} usable configurations (\num{2566} single-chain and \num{12866} ladder), on which all tables below and Figs.~3--4 of the main text rest. Across these \num{15432} configurations, architecture alone explains $R^2=\num{0.211}$ of the variance in $\share$, forward gain $\Gfwd$ alone (\qty{2}{\dB} bins) explains $R^2=\num{0.978}$, adding architecture increases $R^2$ to \num{0.979} ($\Delta R^2=\num{0.0012}$), and $\log_{10}(\mathrm{cond}\,\Sigma)$ explains $R^2=\num{0.959}$. A single
chain of the same length and the same $J,g$ grid provides the baseline. The
comparison is made at fixed $N$, fixed band and fixed $\theta_0$, so that only
the architecture varies.

\begin{table}[htbp]
\caption{Barrier-matched medians. For each band of
$|\Real\nu_{\max}|/\kappa$ and each size: forward gain, reverse gain,
isolation, mean-response share and resource-normalized information of the single
chain (left block) and of the ladder (right block). Gains in \si{\dB}.}
\label{tab:si_matched}
\scriptsize
\setlength{\tabcolsep}{2.5pt}
\begin{ruledtabular}
\begin{tabular}{@{}rrcccccccccc@{}}
$|\Real\nu_{\max}|/\kappa$ & $N$ &
$\Gfwd$ & $\Grev$ & $\Iso$ & $\share$ & $\kappa^2\Fnorm$ &
$\Gfwd$ & $\Grev$ & $\Iso$ & $\share$ & $\kappa^2\Fnorm$ \\
\hline
0.30--0.60 & 8 & \num{39.079} & \num{-169.049} & \num{207.563} & \num{0.075} & \num{105.307} & \num{-28.028} & \num{-82.619} & \num{49.036} & \num{0.935} & \num{1.025}\\
0.30--0.60 & 12 & \num{-32.902} & \num{-187.196} & \num{161.438} & \num{0.937} & \num{1.092} & \num{-58.346} & \num{-111.702} & \num{48.927} & \num{0.953} & \num{1.012}\\
\addlinespace[2pt]
0.20--0.30 & 8 & \num{66.278} & \num{-146.526} & \num{212.804} & \num{4.64e-3} & \num{1825.105} & \num{-4.499} & \num{-54.730} & \num{48.909} & \num{0.787} & \num{1.252}\\
0.20--0.30 & 12 & \num{29.878} & \num{-169.898} & \num{194.384} & \num{0.297} & \num{59.661} & \num{-20.432} & \num{-72.421} & \num{48.684} & \num{0.860} & \num{1.108}\\
\addlinespace[2pt]
0.15--0.20 & 8 & \num{75.371} & \num{-137.433} & \num{212.804} & \num{2.10e-3} & \num{4540.942} & \num{3.503} & \num{-45.210} & \num{48.072} & \num{0.689} & \num{1.740}\\
0.15--0.20 & 12 & \num{63.123} & \num{-174.806} & \num{243.804} & \num{0.038} & \num{6640.997} & \num{-9.580} & \num{-59.197} & \num{47.871} & \num{0.789} & \num{1.275}\\
\addlinespace[2pt]
0.12--0.15 & 8 & \num{81.289} & \num{-131.516} & \num{212.804} & \num{1.38e-3} & \num{8465.636} & \num{7.762} & \num{-40.174} & \num{47.850} & \num{0.601} & \num{2.188}\\
0.12--0.15 & 12 & \num{79.269} & \num{-172.771} & \num{252.041} & \num{0.014} & \num{7e4} & \num{-2.302} & \num{-52.269} & \num{47.986} & \num{0.727} & \num{1.584}\\
\addlinespace[2pt]
0.10--0.12 & 8 & \num{85.625} & \num{-127.179} & \num{212.804} & \num{1.02e-3} & \num{1.3e4} & \num{12.293} & \num{-35.842} & \num{47.982} & \num{0.510} & \num{2.962}\\
0.10--0.12 & 12 & \num{94.081} & \num{-174.432} & \num{268.514} & \num{4.57e-3} & \num{4.7e5} & \num{0.984} & \num{-47.882} & \num{48.937} & \num{0.687} & \num{1.833}\\
\addlinespace[2pt]
0.08--0.10 & 8 & \num{90.470} & \num{-122.334} & \num{212.804} & \num{7.58e-4} & \num{2.1e4} & \num{14.028} & \num{-34.392} & \num{48.036} & \num{0.461} & \num{3.165}\\
0.08--0.10 & 12 & \num{88.972} & \num{-154.832} & \num{243.804} & \num{0.011} & \num{2.6e5} & \num{3.963} & \num{-43.691} & \num{47.615} & \num{0.645} & \num{2.173}\\
\addlinespace[2pt]
0.05--0.08 & 8 & \num{96.213} & \num{-116.591} & \num{212.804} & \num{5.66e-4} & \num{3.7e4} & \num{17.833} & \num{-29.768} & \num{47.636} & \num{0.380} & \num{3.627}\\
0.05--0.08 & 12 & \num{97.400} & \num{-146.404} & \num{243.804} & \num{7.76e-3} & \num{8.6e5} & \num{9.793} & \num{-38.650} & \num{46.252} & \num{0.550} & \num{2.817}\\

\end{tabular}
\end{ruledtabular}
\end{table}

The outcome is stated in the main text and is worth repeating in one line: over
the same distance from threshold the single chain reaches more than five orders of
magnitude more resource-normalized information within one band, and retains two to three orders of magnitude
less of the clean-cavity response in its mean field. Table~\ref{tab:si_gain}
repeats the comparison binned by forward gain rather than by margin, so that the
two families can be read on one scale.

\begin{table}[htbp]
\caption{Medians binned by forward gain, the quantity that tracks the
mean-response share. Both architectures follow one trend, at both sizes and over the whole sampled range of gain; their medians differ by at most \num{0.085}, in the \qtyrange{20}{40}{\dB} band at $N=12$.}
\label{tab:si_gain}
\begin{ruledtabular}
\begin{tabular}{@{}rccccccc@{}}
$\Gfwd$ (\si{\dB}) & $N$ & $\share$ & $\kappa^2\Fnorm$ & $\Iso$ & $\share$ & $\kappa^2\Fnorm$ & $\Iso$ \\
\hline
$\leq 0$ & 8 & \num{0.970} & \num{1.011} & \num{113.216} & \num{0.957} & \num{1.011} & \num{47.962}\\
$\leq 0$ & 12 & \num{0.979} & \num{1.024} & \num{141.804} & \num{0.954} & \num{1.013} & \num{48.171}\\
\addlinespace[2pt]
0--20 & 8 & \num{0.632} & \num{2.975} & \num{134.182} & \num{0.597} & \num{2.203} & \num{52.178}\\
0--20 & 12 & \num{0.671} & \num{4.262} & \num{161.438} & \num{0.598} & \num{2.744} & \num{57.898}\\
\addlinespace[2pt]
20--40 & 8 & \num{0.227} & \num{31.375} & \num{155.148} & \num{0.233} & \num{16.086} & \num{82.488}\\
20--40 & 12 & \num{0.367} & \num{49.223} & \num{177.911} & \num{0.282} & \num{26.834} & \num{89.167}\\
\addlinespace[2pt]
40--60 & 8 & \num{0.035} & \num{338.938} & \num{197.080} & \num{0.036} & \num{71.515} & \num{90.878}\\
40--60 & 12 & \num{0.134} & \num{921.731} & \num{210.857} & \num{0.086} & \num{120.679} & \num{91.105}\\
\addlinespace[2pt]
$\geq 60$ & 8 & \num{2.25e-5} & \num{1.1e4} & \num{291.427} & -- & -- & --\\
$\geq 60$ & 12 & \num{1.03e-3} & \num{1.2e6} & \num{301.460} & -- & -- & --\\

\end{tabular}
\end{ruledtabular}
\vspace{2pt}
{\footnotesize Left block: single chain. Right block: ladder.}
\end{table}

Beyond the medians, the attainable envelope of the box differs sharply between
the families, and Table~\ref{tab:si_envelope} records it. The single chain can be
driven to forward gains above \qty{120}{\dB} at every size and to
\qty{214}{\dB} at $N=10$, with a reverse isolation between
\qty{261}{\dB} at $N=6$ and \qty{469}{\dB} at $N=10$,
while the ladder saturates between \qty{41}{\dB} at $N=12$ and \qty{51}{\dB} at
$N=6$ of forward gain, with \qty{92}{\dB} of isolation at every size. The envelope is the reason the two architectures cannot be compared at
their optima: the ladder is excluded from the region of parameter space where the
chain's information diverges.

\begin{table}[htbp]
\caption{Attainable envelope over the sampled box: the largest resource-normalized
information, its $99$th percentile, the largest forward gain and the largest
isolation reached by each family at each size. Gains in \si{\dB}.}
\label{tab:si_envelope}
\begin{ruledtabular}
\begin{tabular}{@{}rccccc@{}}
$N$ & family & $\kappa^2 F_C^{\rm norm,max}$ & $\kappa^2 F_C^{\rm norm,p99}$ & $G_{\rm fwd}^{\max}$ & $\mathcal{I}^{\max}$ \\
\hline
6 & single & \num{3115.846} & \num{2221.908} & \num{122.291} & \num{260.577}\\
6 & ladder & \num{42.563} & \num{36.809} & \num{51.441} & \num{91.954}\\
8 & single & \num{4.51e5} & \num{2.65e5} & \num{172.215} & \num{364.807}\\
8 & ladder & \num{74.166} & \num{57.185} & \num{46.605} & \num{91.954}\\
10 & single & \num{4.30e7} & \num{1.03e7} & \num{214.281} & \num{469.038}\\
10 & ladder & \num{100.587} & \num{55.254} & \num{42.817} & \num{91.954}\\
12 & single & \num{6.01e8} & \num{2.17e8} & \num{180.566} & \num{441.483}\\
12 & ladder & \num{121.257} & \num{39.623} & \num{40.936} & \num{91.954}\\

\end{tabular}
\end{ruledtabular}
\end{table}

Finally, the fixed-parameter ablation of Tables~\ref{tab:si_ablation} and~\ref{tab:si_ablation_b} holds the
hopping and the non-reciprocity fixed and varies only the inter-chain coupling, pairing only those $(J,g)$ grid points at which both architectures satisfy the stability margin and solver check (which selects toward lower gains as $\lambda$ increases).
At every size and every $\lambda$ the ladder and the chain return the same resource-normalized information, paired at identical $J$ and $g$, to within one percent at the median and six percent at the ninetieth percentile, and the fraction of pairs in which the ladder is the better of the two rises from \num{0.63} to \num{0.99} across the range of $\lambda$ without a change of more than one percent in the median. The inter-chain coupling is the parameter that
moves the operating point along the trade-off curve of Fig.~3 of the main text;
it is not a source of information in itself.

\begin{table}[htbp]
\caption{Fixed-parameter ablation at $N=6$ and $N=8$. At identical $J$ and $g$,
with $\lambda$ varied: number of pairs, median and $90$th percentile of the ratio
$F_C^{\rm norm,ladder}/F_C^{\rm norm,chain}$, and fraction of pairs in which the
ladder is better.}
\label{tab:si_ablation}
\begin{ruledtabular}
\begin{tabular}{@{}rcrccc@{}}
$N$ & $\lambda/\kappa$ & pairs & median ratio & $p90$ ratio & ladder better \\
\hline
6 & \num{5.00e-3} & 572 & \num{1.000} & \num{1.000} & \num{0.633}\\
6 & \num{0.019} & 479 & \num{1.000} & \num{1.004} & \num{0.779}\\
6 & \num{0.032} & 440 & \num{1.001} & \num{1.010} & \num{0.843}\\
6 & \num{0.046} & 412 & \num{1.002} & \num{1.019} & \num{0.891}\\
6 & \num{0.059} & 390 & \num{1.003} & \num{1.031} & \num{0.921}\\
6 & \num{0.073} & 371 & \num{1.005} & \num{1.044} & \num{0.941}\\
6 & \num{0.086} & 357 & \num{1.007} & \num{1.055} & \num{0.952}\\
6 & \num{0.100} & 341 & \num{1.009} & \num{1.060} & \num{0.971}\\
\addlinespace[2pt]
8 & \num{5.00e-3} & 384 & \num{1.000} & \num{1.000} & \num{0.740}\\
8 & \num{0.019} & 334 & \num{1.000} & \num{1.005} & \num{0.874}\\
8 & \num{0.032} & 311 & \num{1.001} & \num{1.015} & \num{0.936}\\
8 & \num{0.046} & 295 & \num{1.002} & \num{1.026} & \num{0.973}\\
8 & \num{0.059} & 284 & \num{1.003} & \num{1.038} & \num{0.986}\\
8 & \num{0.073} & 273 & \num{1.004} & \num{1.041} & \num{0.989}\\
8 & \num{0.086} & 263 & \num{1.006} & \num{1.052} & \num{0.989}\\
8 & \num{0.100} & 254 & \num{1.007} & \num{1.048} & \num{0.992}\\

\end{tabular}
\end{ruledtabular}
\end{table}

\begin{table}[htbp]
\caption{Same as Table~\ref{tab:si_ablation} for $N=10$ and $N=12$.}
\label{tab:si_ablation_b}
\begin{ruledtabular}
\begin{tabular}{@{}rcrccc@{}}
$N$ & $\lambda/\kappa$ & pairs & median ratio & $p90$ ratio & ladder better \\
\hline
10 & \num{5.00e-3} & 271 & \num{1.000} & \num{1.001} & \num{0.804}\\
10 & \num{0.019} & 243 & \num{1.000} & \num{1.007} & \num{0.934}\\
10 & \num{0.032} & 230 & \num{1.001} & \num{1.014} & \num{0.965}\\
10 & \num{0.046} & 219 & \num{1.002} & \num{1.020} & \num{0.982}\\
10 & \num{0.059} & 209 & \num{1.002} & \num{1.021} & \num{0.981}\\
10 & \num{0.073} & 206 & \num{1.004} & \num{1.026} & \num{0.971}\\
10 & \num{0.086} & 198 & \num{1.005} & \num{1.028} & \num{0.985}\\
10 & \num{0.100} & 193 & \num{1.006} & \num{1.030} & \num{0.990}\\
\addlinespace[2pt]
12 & \num{5.00e-3} & 195 & \num{1.000} & \num{1.001} & \num{0.877}\\
12 & \num{0.019} & 178 & \num{1.000} & \num{1.005} & \num{0.961}\\
12 & \num{0.032} & 170 & \num{1.001} & \num{1.007} & \num{0.976}\\
12 & \num{0.046} & 163 & \num{1.001} & \num{1.009} & \num{0.975}\\
12 & \num{0.059} & 160 & \num{1.002} & \num{1.012} & \num{0.988}\\
12 & \num{0.073} & 155 & \num{1.003} & \num{1.014} & \num{0.987}\\
12 & \num{0.086} & 151 & \num{1.005} & \num{1.018} & \num{0.993}\\
12 & \num{0.100} & 148 & \num{1.006} & \num{1.020} & \num{0.993}\\

\end{tabular}
\end{ruledtabular}
\end{table}

\subsection{How the reported intervals are obtained}
\label{sec:si_intervals}

The statistical unit is one usable configuration, that is one point of the box
that survives the stability margin and the solver check
of~\cref{sec:si_reliability}. Two statistics are quoted with intervals. For the
threshold exponents of \cref{sec:si_cuts}, the stable points of a cut at fixed
$g$ are resampled with replacement and the power-law exponent is refitted on each
resample. For the size-resolved share differences, the single-chain and ladder
configurations of one size and one gain band are resampled separately and the
difference of the two medians is recorded. Both use \num{4000} resamples from a
fixed generator seed and quote the \num{2.5} and \num{97.5} percentiles of the
resampled statistic, so the interval describes the variability of the estimate
under resampling of the sampled configurations.

Two cautions apply. No multiplicity correction is made, since the bands and the
cuts were read as a description of the landscape rather than as a family of
pre-registered tests, and an interval that excludes zero is reported as such
rather than as a significance claim. The configurations are also correlated
across size, hopping, coupling and binning, because neighboring points of the
same optimization share their drift and because a cut reuses the same stable
points at successive margins. The intervals should therefore be read as a
descriptive measure of how well the sampled configurations constrain the quoted
statistic, not as the error bar of an independent-sample estimate. For the
architecture comparison of \cref{sec:si_matched} the corresponding statement is
that the two families agree at the median to within one percent and at the
ninetieth percentile to within six percent, computed over the same $\lambda$,
the same four sizes and the paired grid points of the two architectures, on the
resource-normalized information $\Fnorm$ rather than on the raw total $F_C$,
whose two families carry Petermann factors that differ by orders of magnitude.

\section{Threshold cuts}
\label{sec:si_cuts}

The divergence of the single chain's information is a statement about the
approach to the Hurwitz limit, and Table~\ref{tab:si_cuts} makes it explicit by
cutting the box along $J$ at fixed $g$. To avoid eigenvector ill-conditioning of
the spectral solver at small $J$ and large $g$, every cut is evaluated directly
with the Bartels--Stewart Schur solver, retaining points that satisfy
$\max\Real\nu<-0.05\kappa$, relative Lyapunov residual $\le\num{1e-12}$, and
$\mathrm{cond}\,\Sigma\le\num{1e12}$. Ladder curves therefore terminate at their
Hurwitz limit, while strongly non-reciprocal single-chain curves terminate where
$\mathrm{cond}\,\Sigma$ reaches the double-precision ceiling $\num{1e12}$. Over
these validated cuts, the single chain grows as an inverse power of the distance
to threshold with exponents \num{-2.42} at $N=8, g=1$ (\qty{95}{\percent} bootstrap interval
$\numrange{-2.57}{-2.23}$, $R^2=\num{0.93}$), \num{-4.96} at $N=8, g=2$
($\numrange{-5.85}{-4.38}$, $R^2=\num{0.86}$), \num{-6.06} at $N=8, g=2.5$
($\numrange{-7.13}{-5.14}$, $R^2=\num{0.81}$), \num{-3.98} at $N=12, g=1$
($\numrange{-4.26}{-3.59}$, $R^2=\num{0.91}$), \num{-13.10} at $N=12, g=2$
($\numrange{-14.41}{-12.19}$, $R^2=\num{0.95}$), and \num{-23.22} at $N=12, g=2.5$
($\numrange{-26.54}{-20.37}$, $R^2=\num{0.93}$). The ladder curves follow the
single chain closely before destabilizing early, peaking between \num{9.65} and
\num{16.9} at $\lambda=0.02\kappa$ and between \num{1.43} and \num{2.43} at
$\lambda=0.10\kappa$.

\begin{table}[htbp]
\caption{Validated threshold cuts along $J$ at fixed $g$, computed directly with
the Bartels--Stewart Schur solver ($\max\Real\nu<-0.05\kappa$, Lyapunov residual
$\le\num{1e-12}$, $\mathrm{cond}\,\Sigma\le\num{1e12}$): number of retained
points, largest normalized information $\kappa^2\Fnorm$ on the cut, and
power-law fit against distance to the Hurwitz limit.}
\label{tab:si_cuts}
\begin{ruledtabular}
\begin{tabular}{@{}rlcrcccc@{}}
$N$ & family & $\lambda/\kappa$ & $g$ & points & $\kappa^2 F_C^{\rm norm,max}$ & exponent & $R^2$ \\
\hline
8 & single & \num{0.000} & \num{1.000} & 60 & \num{339.061} & \num{-2.420} & \num{0.929}\\
8 & ladder & \num{0.020} & \num{1.000} & 48 & \num{10.482} & \num{-0.901} & \num{0.878}\\
8 & ladder & \num{0.100} & \num{1.000} & 37 & \num{1.765} & \num{-0.223} & \num{0.873}\\
8 & single & \num{0.000} & \num{2.000} & 60 & \num{9.04e4} & \num{-4.962} & \num{0.863}\\
8 & ladder & \num{0.020} & \num{2.000} & 20 & \num{16.562} & \num{-1.488} & \num{0.938}\\
8 & ladder & \num{0.100} & \num{2.000} & 13 & \num{2.341} & \num{-0.446} & \num{0.916}\\
8 & single & \num{0.000} & \num{2.500} & 53 & \num{5.87e4} & \num{-6.062} & \num{0.806}\\
8 & ladder & \num{0.020} & \num{2.500} & 10 & \num{16.924} & \num{-1.775} & \num{0.963}\\
8 & ladder & \num{0.100} & \num{2.500} & 6 & \num{2.434} & \num{-0.557} & \num{0.980}\\
12 & single & \num{0.000} & \num{1.000} & 58 & \num{2.14e4} & \num{-3.981} & \num{0.913}\\
12 & ladder & \num{0.020} & \num{1.000} & 42 & \num{9.649} & \num{-0.706} & \num{0.793}\\
12 & ladder & \num{0.100} & \num{1.000} & 34 & \num{1.586} & \num{-0.156} & \num{0.857}\\
12 & single & \num{0.000} & \num{2.000} & 46 & \num{2.76e6} & \num{-13.103} & \num{0.946}\\
12 & ladder & \num{0.020} & \num{2.000} & 15 & \num{13.872} & \num{-1.278} & \num{0.881}\\
12 & ladder & \num{0.100} & \num{2.000} & 11 & \num{1.887} & \num{-0.302} & \num{0.862}\\
12 & single & \num{0.000} & \num{2.500} & 29 & \num{4.31e5} & \num{-23.224} & \num{0.925}\\
12 & ladder & \num{0.020} & \num{2.500} & 7 & \num{12.662} & \num{-1.642} & \num{0.949}\\
12 & ladder & \num{0.100} & \num{2.500} & 4 & \num{1.431} & \num{-0.355} & \num{0.939}\\

\end{tabular}
\end{ruledtabular}
\end{table}

\section{Control ablations}
\label{sec:si_controls}

Table~\ref{tab:si_controls} reports the control families that isolate the
ingredients of the ladder: the opposing-skin geometry, a parallel-skin geometry
in which both chains accumulate at the same boundary, a decoupled ladder with
$\lambda=0$, and a single chain, optimized over its own hopping and
non-reciprocity at every size. Each family is optimized independently over the
same box and under the same stability barrier. The opposing-skin rows reproduce the ten-restart landscape of~\cref{MAIN-tab:landscape}, while the other three families use a finite-difference optimizer over the same box, barrier, and step budget.

The Petermann factor~\cite{Wang2020Petermann} of the optima reported in the main text is evaluated
separately, from the full biorthogonal eigenbasis of the drift at each converged
operating point. For each mode, $r_i$ is the right eigenvector of $A$ and $\ell_i$
the corresponding left eigenvector of $A^{\dagger}$, both normalized to unit
norm, and $K=1/|\langle\ell_i|r_i\rangle|^{2}$. The spectra involved are not
degenerate, the smallest spacing between distinct eigenvalues being
$\num{0.123}\kappa$ to $\num{0.412}\kappa$, so the pairing between left and right
vectors is unique and no regularization enters. The smallest overlap falls from
$\num{0.024}$ to $\num{0.088}$ over $N=4$ to $16$, giving $K$ between $\num{131}$
and $\num{1800}$; $K=\num{1.000}$ at $N=1$ and $\num{1.46}$ at $N=2$. The
resource-normalized information quoted in~\cref{MAIN-tab:landscape} is not
divided by $K$. One caution applies to this class of estimate. The rows of
$A^{-1}$ returned by a diagonalizing routine are normalized so that
$\ell_i r_i=1$ identically, and an overlap computed as the diagonal of $V^{-1}V$
therefore returns unity for every matrix and every mode, however ill conditioned.
A Petermann factor obtained that way is a tautology rather than a measurement.

The factor also separates the two families, and this is the one place where it
would change an ordering; that is why no ordering is adopted from it. At their converged optima the single chain
gives $K=\num{5e6}$ at $N=4$ rising to $\num{6e26}$ at $N=12$, while the ladder
stays between $\num{131}$ and $\num{1800}$; over the sampled box the median
factor is $\num{4.5e9}$ for the chain and $\num{7.9}$ for the ladder. Dividing
the resource-normalized information by these factors reverses the ordering of
the totals by three to fifteen orders of magnitude. The mean-response share
$\share$ of~\cref{MAIN-eq:share} involves only the mean field and needs no such
convention, which is why it, and not the total, carries the comparison.

The families separate the two ingredients the geometry contributes, and they are
not of equal standing. Removing the link leaves the information bounded, between
$\num{4.0}$ and $\num{42.6}$ at the sizes sampled, and at a given size it differs
from the opposing ladder by a factor of at most thirteen with no systematic sign:
the link is neither what produces the divergence nor what on its own prevents it.
What prevents it is how the link meets the two skins: an excitation amplified rightward along chain A ($\propto e^{+g(N-1)}$) hops across $\lambda$ to chain B, is re-amplified leftward ($\propto e^{+g(N-1)}$), and returns to chain A, creating a closed-loop feedback gain $\propto\lambda^2 e^{2g(N-1)}$ that drives the ladder into Hurwitz self-oscillation. Equivalently, in the frame where each chain is symmetrized, the inter-chain coupling of the opposing geometry is
$\lambda e^{2gj}$, so it grows with the site index and with the non-reciprocity,
and the coupling that still leaves the system inside the barrier falls roughly as
$e^{-4g}$: from $\num{0.214}$ at $g=1$ through $\num{0.055}$ and
$\num{0.0063}$ to $\num{0.0006}$ at $g=2.5$ for $N=6$. A ladder cannot be both
strongly non-reciprocal and coupled, and it is the non-reciprocity that would buy
the gain. This is why the opposing ladder is unstable at the operating point of
the single chain for every $\lambda$ of its box once $N\ge4$, and therefore never
reaches the gain at which the chain's divergence appears. Turning the second
chain around so that both skins accumulate at the same boundary turns the return path into an attenuator ($\propto e^{-g(N-1)}$), removes the exponential loop profile, leaves the coupling uniform, and restores the single-chain growth: $\num{5.7e8}$ at $N=12$ against $\num{12.3}$ for the opposing
geometry, with the single chain at $\num{1.6e9}$ on the same cut. A parallel
ladder is accordingly not a bounded architecture with a different constant but
the chain again, with the same divergence and the same covariance-dominated
budget. One control point, the parallel-skin optimum at $N=8$, settles at
$\Real\nu_{\max}=-\num{0.0096}\kappa$, inside the $0.05\kappa$ margin that
every other family respects; it is quoted only to show where that family goes and
is not a feasible operating point.

The decoupled family is the estimator's own identity check. With
$\lambda=0$ the second chain is undriven, the drift is block diagonal, and the
ladder must return exactly the information of the chain at the same parameters.
Run through the full $2N$-dimensional ladder code it does so to
$10^{-12}$ wherever the covariance is well conditioned, the discrepancy growing
to $10^{-2}$ only at the one configuration whose conditioning reaches $10^{19}$.
The $\lambda=0$ rows of~\cref{tab:si_controls} are nevertheless far below the
chain rows at the same size, and the identity tells us why: the two are different
optima of the same objective, reached from different initializations in a box of
different dimension, not two evaluations of one point. The link is what changes
which optimum exists, and the table is read as a comparison of attractors.

Throughout, the smallest value is attained by the single cavity, which
saturates~\cref{MAIN-eq:cavity} at $N=1$ to within half a percent, as it must.

\begin{table}[htbp]
\caption{Control ablations. Each family is optimized independently over the same
box and under the same barrier; the columns give its optimum and the
resource-normalized information it attains. The opposing-skin rows are the
landscape of~\cref{MAIN-tab:landscape}; the other families use the
finite-difference optimizer described in the text. The inter-chain link is absent
for a single chain and fixed at zero for the decoupled ladder.}
\label{tab:si_controls}
{\small\renewcommand{\arraystretch}{0.88}%
\begin{ruledtabular}
\begin{tabular}{@{}lrccccc@{}}
family & $N$ & $J/\kappa$ & $g$ & $\lambda/\kappa$ & $\kappa^2\Fnorm$ &
$\Real\nu_{\max}/\kappa$ \\
\hline
opposing skin & 1 & -- & -- & \num{0.100} & \num{0.995} & \num{-0.914}\\
opposing skin & 2 & \num{0.474} & \num{0.503} & \num{0.092} & \num{1.126} & \num{-0.429}\\
opposing skin & 4 & \num{0.407} & \num{2.074} & \num{5.009e-3} & \num{12.577} & \num{-0.156}\\
opposing skin & 6 & \num{0.351} & \num{1.729} & \num{5.619e-3} & \num{38.710} & \num{-0.094}\\
opposing skin & 8 & \num{0.293} & \num{1.720} & \num{5.530e-3} & \num{67.198} & \num{-0.070}\\
opposing skin & 12 & \num{0.439} & \num{0.715} & \num{8.573e-3} & \num{12.273} & \num{-0.055}\\
parallel skin & 1 & -- & -- & \num{0.100} & \num{0.995} & \num{-0.913}\\
parallel skin & 2 & \num{0.500} & \num{0.502} & \num{0.100} & \num{1.145} & \num{-0.405}\\
parallel skin & 4 & \num{0.499} & \num{2.998} & \num{0.100} & \num{73.550} & \num{-0.093}\\
parallel skin & 6 & \num{0.092} & \num{2.692} & \num{0.062} & \num{9.672} & \num{-0.773}\\
parallel skin & 8 & \num{0.474} & \num{2.995} & \num{0.100} & \num{5.5e6} & \num{-9.6e-3}\\
parallel skin & 12 & \num{0.445} & \num{2.741} & \num{8.575e-3} & \num{5.7e8} & \num{-0.127}\\
$\lambda=0$ & 1 & -- & -- & \num{0.000} & \num{0.995} & \num{-0.913}\\
$\lambda=0$ & 2 & \num{0.500} & \num{0.502} & \num{0.000} & \num{1.140} & \num{-0.394}\\
$\lambda=0$ & 4 & \num{0.181} & \num{2.572} & \num{0.000} & \num{4.009} & \num{-0.118}\\
$\lambda=0$ & 6 & \num{0.365} & \num{1.699} & \num{0.000} & \num{42.585} & \num{-0.091}\\
$\lambda=0$ & 8 & \num{0.270} & \num{1.468} & \num{0.000} & \num{5.121} & \num{-0.050}\\
$\lambda=0$ & 12 & \num{0.424} & \num{0.719} & \num{0.000} & \num{5.970} & \num{-0.063}\\
single chain & 1 & -- & -- & -- & \num{0.990} & \num{-1.000}\\
single chain & 2 & \num{0.500} & \num{0.502} & -- & \num{1.140} & \num{-0.503}\\
single chain & 4 & \num{0.500} & \num{2.999} & -- & \num{38.267} & \num{-0.193}\\
single chain & 6 & \num{0.500} & \num{2.997} & -- & \num{3104.750} & \num{-0.100}\\
single chain & 8 & \num{0.500} & \num{2.998} & -- & \num{4.5e5} & \num{-0.061}\\
single chain & 12 & \num{0.478} & \num{2.441} & -- & \num{1.6e9} & \num{-0.073}\\

\end{tabular}
\end{ruledtabular}}
\end{table}

\section{Platform parameters and realistic detection}
\label{sec:si_platform}

The model is a linear bosonic lattice written in units of $\kappa$; no mechanical mode enters the drift of~\cref{eq:si_drift}, and the non-reciprocal hoppings are taken as given. To place the dimensionless box on a physical scale we use a membrane-in-the-middle optomechanical array, in which phonon-assisted links of this kind are engineered~\cite{Aspelmeyer2014,Fang2017}, with $\kappa/2\pi=\qty{1.7}{MHz}$, the hopping span of the
sampled box corresponds to $J/2\pi=\qtyrange{85}{850}{kHz}$ and the
inter-chain coupling to $\lambda/2\pi=\qtyrange{8}{170}{kHz}$; a detuning of
$\theta_0=0.1\kappa$ is \qty{170}{kHz}. The non-reciprocity parameter $g$ maps to
a hopping asymmetry difference $\gNR=2J\sinh g = J e^g - J e^{-g}$ between the two hopping directions, which spans
$\gNR/\kappa$ from $0.052$ to $10.0$ over the sampled box and equals $2.65$ at the
reference operating point $J=0.5\kappa$, $g=1.7$. Couplings of this order are within
reach of lithographically defined phonon-assisted links at gaps of a few hundred
nanometers.

The detection budget is the second half of the model. For $N\ge2$, end-to-end transmission along chain A is evaluated from $S_{21}=-\kappa_{\rm port}(A^{-1})_{N-1,0}$ with symmetric port coupling $\kappa_{\rm port}=\sqrt{2}\kappa\le 2\kappa$, so the forward power gain is $G_{\rm fwd}(\si{\dB})=10\log_{10}|S_{21}|^{2}$ and the linear power transmission is $T=|S_{21}|^{2}=10^{G_{\rm fwd}(\si{\dB})/10}$. As an order-of-magnitude benchmark on the displacement channel, if a downstream detection chain adds $n_{\rm add}$ quanta of phase-insensitive noise~\cite{Clerk2010,Lau2018}, the output signal-to-noise ratio is diluted by $T/(T+2n_{\rm add})$ (a full detector-propagated output Fisher calculation is outside the intracavity model of \eqref{eq:si_split}). Across the ladder optima for $N\ge2$, the forward gain is positive everywhere, from $+\qty{3.4}{\dB}$ at $N=2$ ($T=\num{2.2}$) to $+\qty{41.6}{\dB}$ at $N=8$ ($T=\num{1.4e4}$). For $n_{\rm add}=5$ this dilution factor rises from \num{0.18} at $N=2$ to \num{0.9993} at $N=8$, so downstream added noise is not the binding constraint on the amplifying optima; what binds is the distance to the stability threshold, through the covariance term of \eqref{eq:si_split}.

Added noise also acts as a state-level robustness test when added directly to the covariance, $\Sigma\to\Sigma+n_{\rm add}I$. For an uncoupled single cavity ($\Sigma=I$), isotropic noise scales the mean-response share exactly as $\share=1/(1+n_{\rm add})$, falling from \num{1.000} to \num{0.500} at $n_{\rm add}=1$ and \num{0.167} at $n_{\rm add}=5$. At the $N=1$ coupled ladder optimum ($\lambda=0.0998\kappa$), the share similarly drops from \num{1.000} to \num{0.504} for one added quantum and to \num{0.169} for five, whereas at the amplifying $N=4$ single-chain optimum the ratio $\Fcov/\Fmean$ moves only from \num{4.6e5} to \num{4.4e5} for one quantum and \num{4.1e5} for five. A configuration whose information sits in the covariance is therefore not displaced from that regime by isotropic noise.

This is also where the mean-response share earns its role. The covariance term is
the derivative of the noise with respect to the signal, and measuring it requires
resolving a change in the noise against the noise itself; any phase-insensitive
noise added by the detection chain enters that comparison directly, whereas the
mean-response term is a displacement of the field and is limited only by the
transmittance. The two terms are therefore not interchangeable at fixed photon
count, and the split \eqref{eq:si_split} is the minimal accounting that keeps
them apart.

\section{Limitations}
\label{sec:si_limitations}

Three limitations bound the scope of the conclusions.

The dynamics are linearized and Gaussian. This is the regime in which the
Gaussian Fisher information has a closed form and in which the architecture was
proposed, but it excludes the nonlinear and non-Gaussian resources that a
strongly driven optomechanical system can provide, and it excludes photon
counting, which carries information that a heterodyne record does not.

The detection model is a dilution. We account for the added noise of the
detection chain through the transmittance factor $T/(T+2n_{\rm add})$ and do not
model a specific amplifier, filter or feedback loop. A phase-sensitive
amplification scheme would change the relative weight of the two terms of
\eqref{eq:si_split}, and the mean-response share is the quantity that would
change first. The isotropic test of Sec.~IX is a robustness check on the state
rather than a detection model: it asks whether the split survives noise added
directly to the covariance, not how a given detector records the field.

The sampled box is finite. The conclusion that the share is a function of the
forward gain alone is established over $J=0.05$--$0.5\kappa$,
$g=0.5$--$3.0$ and $\lambda=0.005$--$0.1\kappa$ at $N=6$--$12$, with the
landscape of Table~\ref{tab:si_landscape} extending to $N=1$ and $N=16$. We have
no evidence of a different behavior outside that box. The relation is an empirical statement over that box; the one case in which it is exact is the single cavity, whose covariance carries no parameter dependence and whose resource-normalized information is $\kappa^2\Fnorm=1/[1+(\theta/\kappa)^{2}]$ in closed form.

\bibliographystyle{unsrtnat}
\bibliography{04_track_C_nhse_references}